\documentclass[nofootinbib,amsmath,twocolumn,notitlepage,preprintnumbers]{revtex4-1}
\usepackage{multirow}
\usepackage{amssymb, esvect, amsmath, graphicx, latexsym, amsthm, slashed, eso-pic}
\usepackage{xcolor}
\usepackage{amsmath,amsthm,amsfonts,amscd,amssymb} 
\usepackage{eucal}
\usepackage{braket}
\usepackage{slashed}
\usepackage{hyperref}
\usepackage{graphicx}
\usepackage{bm}
\usepackage{siunitx}
\usepackage{xcolor}
\usepackage{afterpage}
\usepackage{changepage}

\usepackage{hyperref}

\newcommand{\beq}{\begin{equation}} \newcommand{\eeq}{\end{equation}}
\newcommand{\bea}{\begin{eqnarray}} \newcommand{\eea}{\end{eqnarray}}

\newcommand{\Oi}{\mathcal{O}}

\def\lsim{\mathrel{\raise.3ex\hbox{$<$\kern-.75em\lower1ex\hbox{$\sim$}}}}
\def\gsim{\mathrel{\raise.3ex\hbox{$>$\kern-.75em\lower1ex\hbox{$\sim$}}}}

\newcommand{\be}{\begin{eqnarray}}
\newcommand{\ee}{\end{eqnarray}}

\newcommand{\benum}{\begin{enumerate}}
\newcommand{\eenum}{\end{enumerate}}
\newcommand{\bi}{\begin{itemize}}
\newcommand{\ei}{\end{itemize}}

\begin{document}

\preprint{FERMILAB-PUB-19-537-T}

\title{Back-action evading impulse measurement with mechanical quantum sensors}

\author{Sohitri Ghosh$^{a,b}$}
\thanks{sohitri@umd.edu}
\author{Daniel Carney$^{a,b,d}$}
\thanks{carney@umd.edu}
\author{Peter Shawhan$^{b,c}$}
\thanks{pshawhan@umd.edu}
\author{Jacob M. Taylor$^{a,b}$}
\thanks{jmtaylor@umd.edu}

\medskip

\affiliation{$^a$ Joint Quantum Institute/Joint Center for Quantum Information and Computer Science, University of Maryland, College Park/National Institute of Standards and Technology, Gaithersburg, MD, USA}
\affiliation{$^b$ Department of Physics, University of Maryland, College Park, MD, USA}
\affiliation{$^c$ Joint Space-Science Institute, University of Maryland, College Park/NASA Goddard Space Flight Center, Greenbelt, MD, USA}
\affiliation{$^d$ Fermi National Accelerator Laboratory, Batavia, IL, USA}

\date{\today}

\begin{abstract}
The quantum measurement of any observable naturally leads to noise added by the act of measurement. Approaches to evade or reduce this noise can lead to substantial improvements in a wide variety of sensors, from laser interferometers to precision magnetometers and more. In this paper, we develop a measurement protocol based upon pioneering work by the gravitational wave community which allows for reduction of added noise from measurement by coupling an optical field to the momentum of a small mirror. As a specific implementation, we present a continuous measurement protocol using a double-ring optomechanical cavity. We demonstrate that with experimentally-relevant parameters, this protocol can lead to significant back-action noise evasion, yielding measurement noise below the standard quantum limit over many decades of frequency.
\end{abstract}

\maketitle

\section{Introduction}

 The precision of any measurement is limited by noise. Beyond technical sources of noise such as thermal backgrounds, quantum mechanics imposes a fundamental source of noise: the act of measurement itself can disturb the system being observed. However, the noise added by the measurement depends on how and what we probe, and in some settings can be reduced or even removed by a judicious choice of measurement protocol.
 
 In an optomechanical system such as a laser interferometer, the noise added by measurement can be decomposed into two parts. The first is shot noise, coming from the finite counting statistics of the photons used to probe the mechanical system. The other is measurement back-action noise, which arises because of fluctuations in the radiation pressure of the light \cite{caves1980quantum,caves1981quantum}. Early on, it was realized that direct momentum measurements could be used to reduce measurement-added noise \cite{braginsky1980quantum,caves1980measurement}. In the context of gravitational wave detection, Braginsky and Khalili proposed a concrete velocity-meter scheme in 1990 \cite{braginsky1990gravitational}, and this idea has recently been revisited  \cite{braginsky2000dual,purdue2002practical,chen2003sagnac,danilishin2004sensitivity,danilishin2018new,danilishin2019advanced}, with a prototype experiment in progress \cite{graf2014design}. Other recent work has proposed using a discrete momentum measurement for noise reduction in sensing of forces \cite{khosla2017quantum}.

Our approach here is to  examine the use of continuous momentum measurement to evade back-action noise in the setting of broadband force sensing, i.e., the detection of rapid impulses. For other approaches to back-action evasion, see for example \cite{pereira1994backaction,clerk2008back,hertzberg2010back,woolley2013two,vasilakis2015generation}. We present a treatment from a purely quantum optics perspective to demonstrate the benefits of a ``speedmeter" design  for application in a wide variety of sensors beyond gravitational wave detection. The core idea of our scheme is to monitor the momentum of a mechanical system by coherently integrating the discrete time derivative of the position. In the limit of low optical losses, this becomes equivalent to direct momentum measurement. In the ideal setting of a free-falling mirror without dissipation or losses, such a measurement would enable a complete elimination of all measurement-added noise, since one can further eliminate shot noise by ramping up the probe laser power. To examine imperfections and experimental challenges, we study a practical implementation using a pair of ring cavities, including loss and mechanical noise, which still allows for significant reduction of measurement-added noise. 

The detection of rapid, small impulses is ubiquitous in physics, and these ideas should have broad applicability. In metrology, our broadband approach for optomechanical sensing enables applications such as detection of individual low-energy photons or gas collisions with a mechanical element, which would enable quantum noise-limited pressure calibrations \cite{fremerey1982spinning,looney1993measurement,hodges1996laser,arpornthip2012vacuum,makhalov2016primary,eckel2018challenges} and force sensing  \cite{tsang2010coherent,melcher2014self,xu2014squeezing,wimmer2014coherent,motazedifard2016force}. In particle physics, low-threshold detection of energy deposition is of crucial importance in many contexts, for example the detection of light dark matter candidates \cite{essig2016direct,schutz2016detectability,essig2017detection,knapen2018detection} and astrophysical neutrinos \cite{krauss1991low,mckinsey2000liquid,bellini2014final,ranucci2016techniques}. A concrete application which drove this work is the detection of tiny gravitational forces generated by transient dark matter particles \cite{Carney:2019pza}, and this example is studied in detail in section \ref{longrange}.

\section{Continuous momentum measurement}

We begin with a conceptual outline of the advantages that momentum sensing can provide over position sensing in the context of short signals. Consider the classic argument for the ``standard quantum limit'' (SQL) in a position measurement \cite{caves1980quantum,caves1981quantum}. By measuring the system's position, we reduce the position uncertainty $\Delta x$ while increasing its momentum uncertainty $\Delta p$. Assuming the system Hamiltonian is essentially free between measurements, this state will spread to an uncertainty $\Delta x' = \Delta x + \tau \Delta p/m$ after a short time $\tau$. Thus, a subsequent measurement of the system will suffer from this increased uncertainty. One could try to probe the system with more measurements (decrease the shot noise), but this will increase the momentum spread $\Delta p$ (increase the back-action). Position measurement then involves a fundamental trade-off between these two effects; the optimization leads to the SQL uncertainty $\Delta x^2_{SQL} = \hbar \tau/m $.

Momentum measurement, on the other hand, does not suffer from this competition. If we first measure the momentum of the system, this will decrease the momentum uncertainty $\Delta p$. The subsequent free evolution of the system will then preserve this uncertainty, since $[H,p] = 0$. One can therefore monitor the momentum with arbitrarily low noise by increasing the rate or strength of these momentum measurements. This represents the ``quantum non-demolition'' nature of the momentum measurement \cite{braginsky1980quantum}. If the measuring apparatus is not truly free but has an external potential, for example a harmonic trap, the non-demolition behavior should hold as long as we perform measurements much faster than the internal dynamics of the device.

\begin{figure}[t!]
\includegraphics[scale=1]{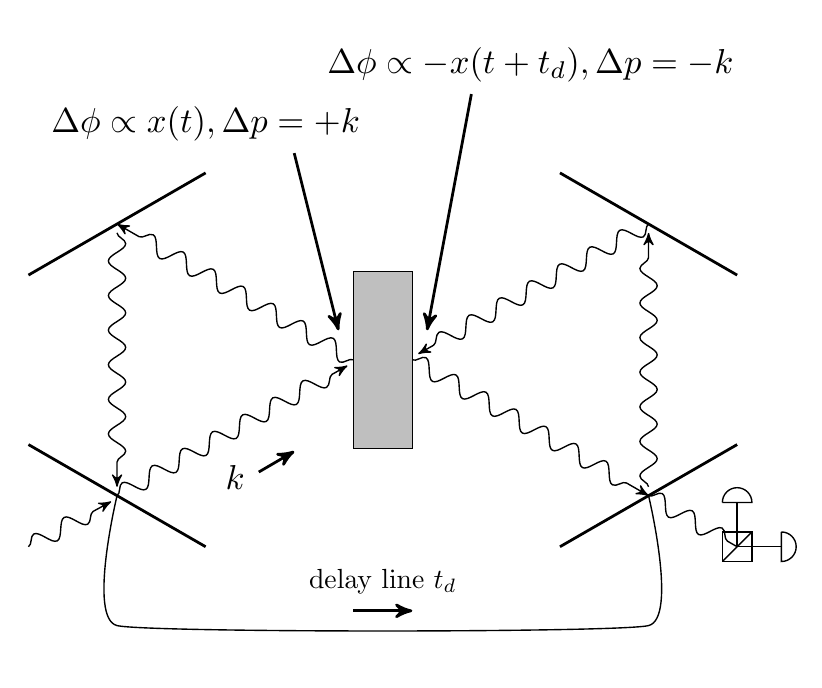} \\
\includegraphics[scale=0.47]{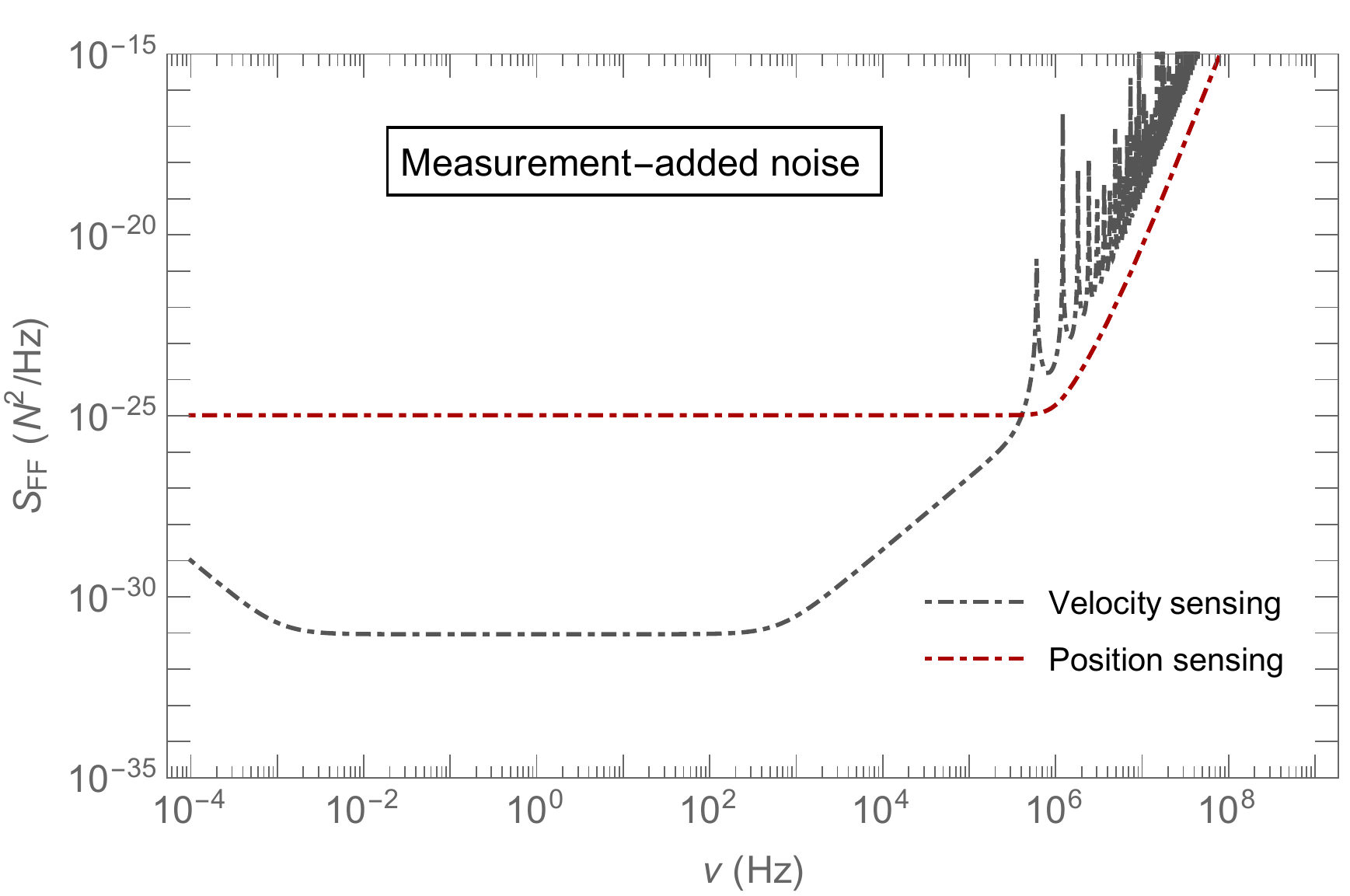}
\caption{Top: Concrete realization of a velocity measurement, using a pair of optical ring cavities separated by a delay line with a suspended mirror as the detector. Each probe photon imparts a momentum $+k$ on the central mirror in the first cavity. After going through a short delay line and into the second cavity, the photon then imparts a momentum $-k$ on the mirror, and is finally read out by an interferometer. The net phase picked up by the light is $\Delta \phi \propto x(t) - x(t + t_d) \propto v$. The two impulses cancel and lead to zero net impulse on the mirror, which amounts to a back-action evading measurement. Bottom: Quantum measurement-added force noise (shot noise plus back-action) in typical position sensing and velocity sensing protocols. Reduction of the back-action noise in the velocity sensing protocol leads to substantial improvements in force sensitivity over a broad band at low frequencies. See figure \ref{figure-noisecontribs} for detailed explanations of the behavior in this plot as well as the relevant detector parameters.}
\label{fig-result}
\end{figure}

To illustrate this general idea, we now study a concrete optomechanical realization where two ring cavities share a common mechanical element, a two-sided mirror (see figure \ref{fig-result}). The light interacts with the shared mirror twice from opposite directions with a short time delay $t_d$. In the first interaction, the light picks up a phase shift proportional to the mirror position $x(t)$ at the time of interaction $t$. After being run through an optical delay line and fed into a second cavity, the same light then picks up an additional phase shift $\propto -x(t + t_d)$, the mechanical position at the time of the second interaction. This imprints a discrete estimate of the mechanical velocity onto the phase of the light $\phi \sim x(t)-x(t + t_d)$, which can then be read out directly through an interferometer. The same basic setup was proposed in \cite{braginsky1990gravitational,danilishin2018new} as a ``speedmeter'', with the goal of searching for gravitational waves. Here we focus instead on the use of this protocol for direct sensing of small impulses (momentum transfers) on a mechanical element.

From this picture, one can see the microscopic mechanism for evasion of the back-action noise: fluctuations in the laser radiation pressure are equal and opposite between the two subsequent light-mechanical interactions, leading to a total change in the mirror momentum that approaches zero. In a practical setting, this cancellation is limited by optical losses. We now study this model using the tools of quantum optics to understand the roles of imperfections and noise in limiting this system for momentum measurement.

We remark that a short, sharp force applied to the system leads to a breaking of the quantum non-demolition condition. Thus, measuring the effect of a sharp force over a short time would best be done by monitoring the momentum before and after the event, such as a gas molecule hitting the mirror. However, here we show that continuous momentum measurement provides a similar benefit in the free fall limit.

\subsection{Detector configuration and noise}

Consider a pair of optical ring cavities which share a common mechanical element, taken to be a harmonic oscillator (e.g.\ a high-quality mirror suspended as a pendulum) with natural frequency $\omega_m$, used here as a resonator. If we monitor the system much more rapidly than its mechanical frequency, we can approximate the dynamics as those of a freely falling system with $\omega_m \to 0$. While $\omega_m \rightarrow 0$ can be achieved by simply dropping the system, keeping a mechanical tether and thus a finite $\omega_m$ allows us to track corrections from a confining potential to the quantum non-demolition benefits that we can hope to realize.

The combined optomechanical system formed by the cavities, mechanical resonator, and their baths can be characterized by the Hamiltonian
\begin{align}
\begin{split}
H_{\text{tot}} &= H_{\rm{cav}}  +H_{\rm{mech}} + H_{\text{bath}}\\
H_{\rm{cav}} &= \hbar \omega a^\dagger a +  \hbar \omega^\prime  a^{\prime\dagger} a^\prime\\
H_{\rm{mech}} &= \frac{p^2}{2m}+ \frac{1}{2}m \omega_m^2 x^2.
\end{split}
\end{align}
Here $a$ and $a^\prime$ denote the annihilation operators for the optical cavities. The frequencies $\omega$ and $\omega^\prime$ are the resonance frequencies of the cavities. These are functions of the lengths of the cavities, which in turn depend on the mechanical displacement $x$ of the resonator: $\omega = \omega(x), \omega' = \omega'(x)$. For small displacements, we can Taylor expand the frequencies, and obtain
\begin{align}
\begin{split}
H_{\text{sys}}&= \hbar \omega_c  \left( a^\dagger a +  a^{\prime\dagger} a^\prime\right) + H_{\rm{mech}} \\
& + \hbar  \left( g_0 a^\dagger a + g_0^\prime a^{\prime\dagger} a^\prime\right) \frac{x}{x_0}.
\end{split}
\end{align}
The first line contains the kinetic terms for the two cavities and the resonator, where we have taken the two cavities to have the same frequency $\omega_c = \omega(0) =\omega^\prime (0)$ when the mirror is at its equilibrium position. The second line encodes the optomechanical coupling with strength $g_0 = x_0 \frac{d\omega}{dx} = -x_0 \omega_c/\ell = -g_0'$, where  $\ell$ is the equilibrium length of the cavity and $x_0 $ is a length parameter which transforms the coupling strength to a frequency. The key point is that the two cavity-mechanical couplings differ by a relative minus sign, corresponding to the fact that displacements of the resonator generate opposite frequency shifts in the two cavities.

We now examine the system using the input-operator formalism \cite{gardiner1985input} (see appendix \ref{appendix-singlesided} for a review in the case of a single-sided cavity). To understand the measurement procedure, we consider driving the first cavity with a monochromatic laser. This effectively displaces the cavity operators by $a\rightarrow (\alpha+a )e^{-i \omega_L t}$, with $\omega_L$ being the frequency of the monochromatic laser and $\alpha \propto \sqrt{P/\hbar \omega_L \kappa}$ being the drive strength in terms of the laser power $P$ and cavity energy loss rate $\kappa$. We have factored out the drive-frequency time dependence in the light fluctuations (i.e.\ we work in the frame co-rotating with the drive). We assume sufficient driving $|\alpha| \gg 1$ so that we can linearize the interaction Hamiltonian around the drive. We choose a gauge such that $\alpha$ is purely real, and lock the laser to provide zero detuning $\Delta = \omega_L - \omega_c = 0$. We then obtain the total Hamiltonian for the system
\begin{equation}
\begin{split}
H_{\text{sys}} = H_{\rm{mech}}
+\hbar G x X -\hbar G^\prime x X^{\prime}.
\end{split}
\end{equation}
Here $X = (a+a^{\dagger})/\sqrt{2}$ and similarly $X'$ are the amplitude quadratures of the cavity modes. The drive enhances the effective optomechanical coupling strength $G= \sqrt{2}\frac{g_0}{x_0} \alpha $ in the first cavity which has the dimension of a frequency per length; in the second cavity, we have $G^\prime =-\sqrt{1-L}G$, where $L$ represents loss of photons as they traverse the delay line. We will justify this shortly [see \eqref{input-output}].

The system is subject to dissipation via the bath $H_{\rm bath}$. The cavity bath consists of the cavity photons leaking through the mirrors and the mechanical bath consists of, at least, ambient gas molecules in the chamber and phonons in any support structure. Tracing out the bath with the input-operator formalism, we can study the evolution of the system in the Heisenberg picture. The equations of motion of the cavities and resonator are 
\begin{align}
\begin{split}
\label{EOM}
\dot{X} &= -\frac{\kappa}{2}X+ \sqrt{\kappa} X_{in}\\
\dot{Y} &= -G x -\frac{\kappa}{2}Y+ \sqrt{\kappa} Y_{in}\\
\dot{X^{\prime}} &= -\frac{\kappa}{2}X^{\prime}+ \sqrt{\kappa} X_{in}^{\prime}\\
\dot{Y^{\prime}} &= G^\prime  x -\frac{\kappa}{2}Y^{\prime}+ \sqrt{\kappa} Y_{in}^{\prime}\\
\dot{p} &= -\hbar G X +\hbar G^\prime  X^{\prime} - m\omega_m^2 x -  \gamma p +  F_{in}\\ 
\dot{x} &= \frac{p}{m}.
\end{split}
\end{align}
Here, $\gamma$ is the mechanical energy damping rate, $F_{in}$ is the external force (including noise) incident on the resonator, $Y = -i(a-a^{\dagger})/\sqrt{2}$ and similarly $Y'$ are the phase quadratures of the cavities, and $X_{in}, Y_{in}, X'_{in},Y'_{in}$ represent the vacuum fluctuations of the cavities. These satisfy the white noise correlation functions of the form
\begin{align}
\begin{split}
\label{vacuum}
\braket{X_{in}(t) X_{in}(t')} & = \braket{Y_{in}(t) Y_{in}(t')} = \delta(t-t') \\
\braket{X_{in}(t) Y_{in}(t')} & = 0,
\end{split}
\end{align}
and similarly for the primed correlators. The force noise will be discussed in detail when needed.

Each cavity has both an input and output field associated to it. Ultimately we want to read out the phase quadrature of the second cavity $Y'_{out}$. The output fields are related to the input fields by the usual input-output relations (again see appendix \ref{appendix-singlesided} for a review). Here we also need to model the delay line. Photons traversing the delay line can be lost, so we model the line as a beam splitter with a dimensionless loss coefficient $L$, leading to the input-output relations
\begin{align}
\begin{split}
\label{input-output}
X_{\text{out}}(t) &= X_{\text{in}}(t) - \sqrt{\kappa} X(t)\\
Y_{\text{out}}(t) &= Y_{\text{in}}(t) - \sqrt{\kappa} Y(t)\\
X^{\prime}_{\text{out}}(t) &= X^{\prime}_{\text{in}}(t) - \sqrt{\kappa} X^{\prime}(t)\\
Y^{\prime}_{\text{out}}(t) &= Y^{\prime}_{\text{in}}(t) - \sqrt{\kappa} Y^{\prime}(t) \\
X^{\prime}_{\text{in}}(t) &= \sqrt{1-L} X_{\text{out}}(t-t_d) + \sqrt{L}\tilde{X}_{\text{in}}(t)\\
Y^{\prime}_{\text{in}}(t) &= \sqrt{1-L} Y_{\text{out}}(t-t_d) + \sqrt{L}\tilde{Y}_{\text{in}}(t).
\end{split}
\end{align}
Here  $\tilde{X}_{\text{in}} , \tilde{Y}_{\text{in}}$ are the input noise fields associated with the loss in the delay line, taken again to satisfy the vacuum noise correlations \eqref{vacuum}. The last two equations here justify the relation $G' = -\sqrt{1 - L}G$ between the two driven coupling strengths.

We are interested in monitoring the external force $F_{in}$ acting on the mechanical system. This force is imprinted onto the mechanical displacement $x(t)$. Working in the frequency domain, we can easily solve the equations of motion \eqref{EOM}, \eqref{input-output} to find the mechanical displacement:
\be
x(\nu) = \chi_m(\nu) F_{\text{in}}(\nu)  + x_{\rm n}(\nu),
\ee 
where the term due to measurement noise is
\begin{equation}
\begin{split}
x_{\rm n} &=-
\hbar G \chi_m \chi_c \left[ \left( 1 + (1-L) e^{i(\nu t_d+\phi_c)} \right) X_{\text{in}}\right. \\
& \left. + \sqrt{L(1-L)}\tilde{X}_{\text{in}}\right].
\end{split}
\end{equation}
Here we defined the cavity and the mechanical response functions and the phase,
 \be
 \begin{split}
 \label{responsefns}
 \chi_c = \frac{\sqrt{\kappa}}{-i \nu+\kappa/2} \; \; \; \; \; 
 & \chi_m = \frac{-1}{m(\nu^2-\omega_m^2 +i\gamma \nu)} \\
 e^{i\phi_c} &= 1-\sqrt{\kappa} \chi_c . 
 \end{split}
 \ee
 At very low frequency, $\nu \approx 0$ and with a small amount of loss $L\approx 0$, $G^\prime \rightarrow -G$ and $e^{i \phi_c} \rightarrow -1$, thus the term proportional to the input noise $X_{\rm in}$ in the position variable vanishes. This amounts to back-action evasion in the low frequency part of the measurement: there is no net force from the fluctuations in the radiation pressure. The noise from the delay loss $\tilde{X}_{in}$ will also be negligible in this limit.
 
The mechanical displacement is in turn imprinted onto the phase quadrature $Y$ of the light through the optomechanical coupling $H_{\rm int} \sim G x X$. We then read out the output light $Y'_{out}$ from the second cavity, from which we infer the external force $F_{in}$. The equations of motion \eqref{EOM}, \eqref{input-output} yield the output light phase
\begin{equation}
\begin{split}
\label{Yout}
Y^{\prime}_{\text{out}}
&= \sqrt{L} e^{i \phi_c} \tilde{Y}_{\text{in}} + \sqrt{1-L} e^{i(\nu t_d+2\phi_c)}  Y_{\text{in}} \\
& + G \sqrt{1-L} \chi_c \left( 1 + e^{i (\nu t_d + \phi_c)} \right) x.
\end{split}
\end{equation}
Given the measured output light $Y'_{out}$, we estimate the force by simply dividing through with the appropriate coefficient:
\begin{equation}
\label{FE}
F_E = \frac{ Y'_{out} }{G \sqrt{1-L}  \chi_c \chi_m \left(1+e^{i(\nu t_d+\phi_c)}\right)}.
\end{equation}

In order to calculate our sensitivity to various signals, we need the noise in the force estimator. We define the force noise power spectral density (PSD) in the usual way,
\begin{equation}
\label{FEPSD}
 \braket{F_E(\nu)F_E(\nu^\prime)} = N(\nu) \delta(\nu+\nu^\prime)= S_{FF}(\nu) \delta(\nu+\nu^\prime).
\end{equation}

\begin{figure}[t]
\includegraphics[scale=0.47]{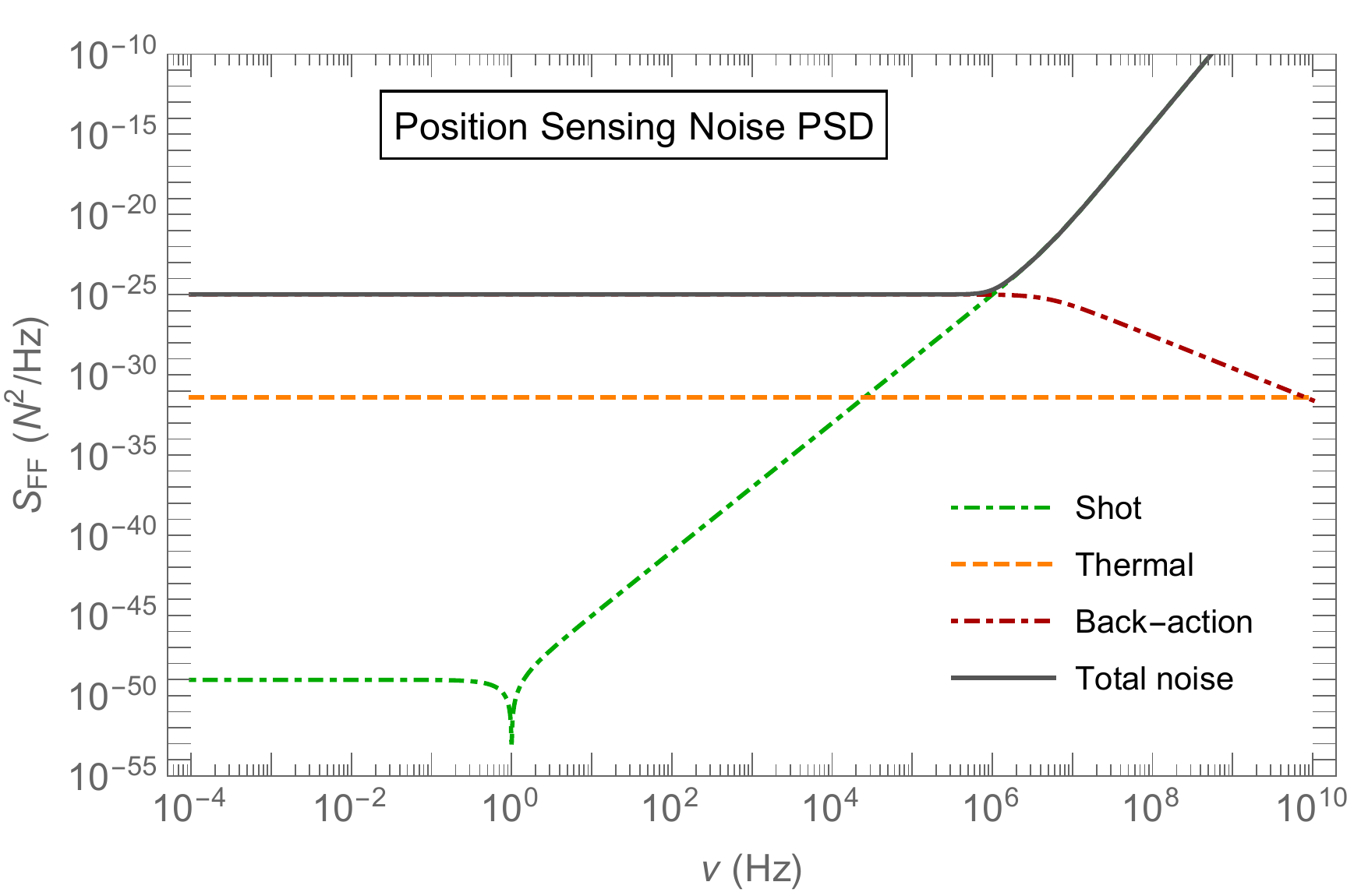} \\
\includegraphics[scale=.47]{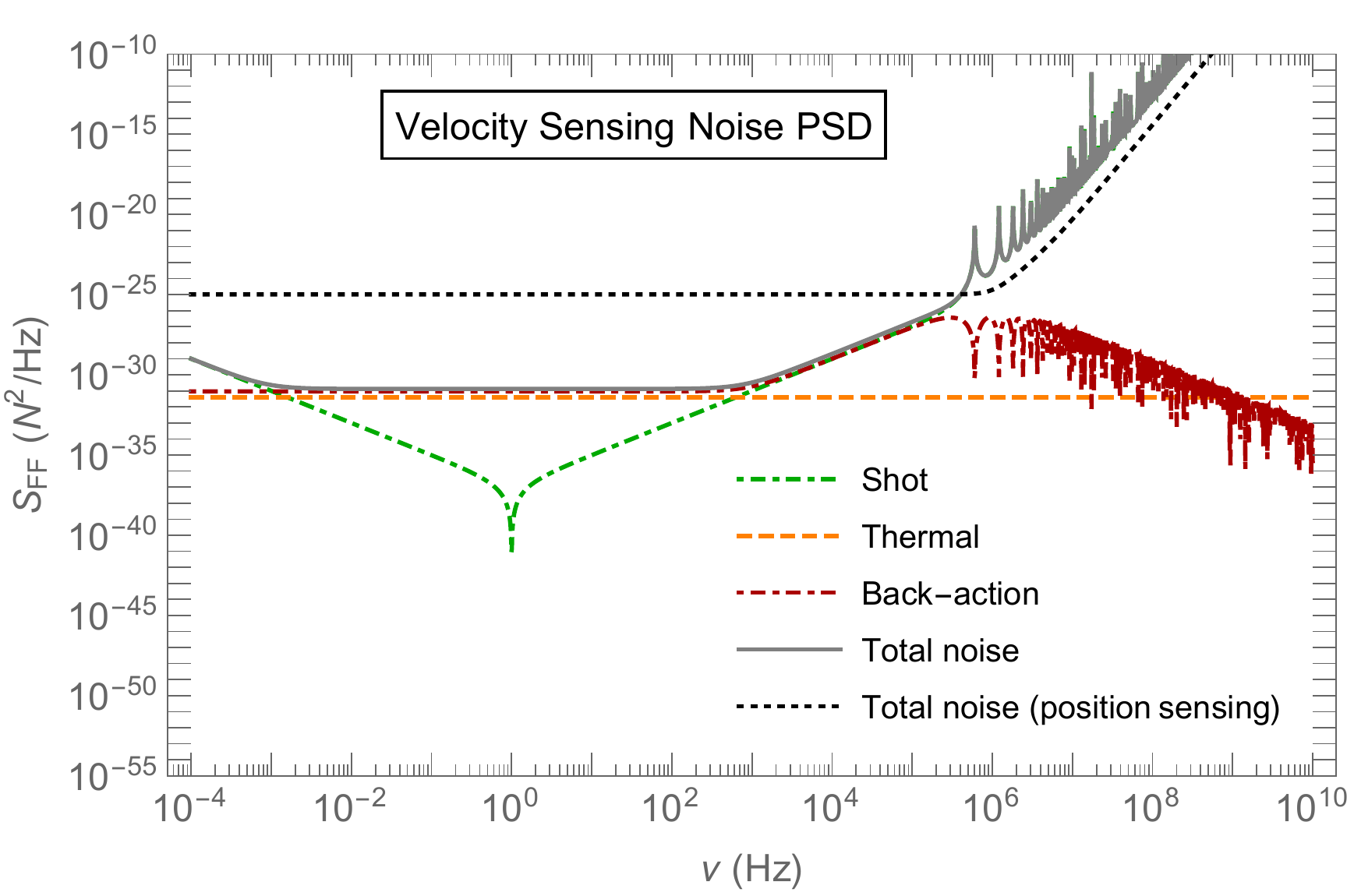}
\caption{Force noise power spectral densities for position sensing in a standard single-sided optomechanical cavity (top) and our velocity sensing protocol in a double ring cavity (bottom). See equations \eqref{noisepsdsingle} and \eqref{noisepsd}, respectively. In addition to the spectrally flat thermal noise, we have shot noise and measurement back-action curves. One can clearly see that back-action noise in the velocity sensing protocol is substantially reduced in the range $\omega_m \lesssim \nu \lesssim \kappa$. The spiky features at high frequency come from resonance effects in the two-cavity response function $\propto e^{i [\nu t_d + \phi_c(\nu)]}$. Here in both cases the detector parameters are taken as $m=1 ~ \rm{g}, \omega_m = 1 ~\rm{Hz}, \gamma = 10^{-4} ~\rm{Hz}, \kappa = 10 ~\rm{MHz}$, at bath temperature $T = 10  ~\rm{mK}$. In the position sensing protocol, optomechanical coupling $G$ is optimized as in equation \eqref{optimizedGsingle} with $\tau = 1~\rm{\mu s}$. In the velocity sensing protocol, the delay line parameters are taken as $t_d= 10 ~\rm{\mu s}, L = 10^{-4}$, and the coupling is optimized as in \eqref{optimizedG}.}
\label{figure-noisecontribs}
\end{figure}

We will see later how exactly this is used to determine sensitivities, but the intuition is that for a broadband impulse signal, sensitivities are set by an integral of $N(\nu)$ over the relevant frequency band. Let us assume that the input force $F_{in}$ is purely thermal (Johnson-Nyquist) noise, so that
\begin{align}
\label{thermalnoise}
\braket{F_{in}(t) F_{in}(t')} = N_{BM}  \delta(t-t'), \ \ N_{BM} = 4 m \gamma k_B T
\end{align}
with $T$ the temperature of the bath coupled to the resonator. Then the
force noise PSD can be computed directly using \eqref{Yout}, \eqref{FE}, \eqref{FEPSD} and the vacuum noise correlation functions \eqref{vacuum}:
\begin{equation}
\begin{split}
\label{noisepsd}
N(\nu) &=\frac{ 1}{4  (1-L) |G |^2 |\chi_c|^2 |\chi_m|^2 \cos^2(\frac{\nu t_d+\phi_c}{2})} \\
& +  N_{BM} + 2   \hbar^2 |G|^2  |\chi_c|^2 \left[1-\frac{L}{2} \right.\\
& \left. +\frac{(1-L)}{\nu^2+\kappa^2/4} [\nu \kappa \sin(\nu t_d) +(\nu^2 -\kappa^2/4) \cos(\nu t_d)]  \right].
\end{split}
\end{equation}
The first term here is the shot noise arising from the statistical counting errors of the laser photons. The middle term is the thermal noise.  The last term denotes the back-action noise arising from the light pushing the mirror around while probing the system. At low frequency, the term in the bracket is proportional to $L/2$ and thus the back-action noise vanishes to the lowest order of the loss coefficient. We plot this noise PSD in figure \ref{figure-noisecontribs}. For comparison, we present the analogous noise PSD for a standard single-sided cavity force sensor \eqref{noisepsdsingle} in the same figure (see appendix \ref{appendix-singlesided} for details).

In order to get some intuition about the noise in this protocol, first consider the high- and low-frequency behavior of the noise PSD \eqref{noisepsd}. At arbitrarily low frequencies $\nu$, the noise PSD diverges:
\be
N(\nu) \xrightarrow[\nu \to 0]{} \frac{ m^2 \kappa^3 \omega_m^4}{4 |G|^2 (4 +\kappa t_d)^2}\frac{1}{\nu^2} \ \ \ .
\ee
Meanwhile, at very high frequencies we also have a divergence
\be
N(\nu) \xrightarrow[\nu \to \infty]{} \frac{ m^2}{ |G|^2 \kappa } \nu^6 \ \ \ .
\ee
These two limits mean that for a given broadband force signal, the very low and very high frequency parts of the signal will not be visible above our noise. Thus our protocol automatically comes with an effective bandwidth. We will discuss this in detail in examples. 

We will be particularly interested in signals of very short temporal duration $\tau$. If this is short compared to the natural period of the mechanical resonator, $\omega_m \tau \ll 1$, then the resonator is essentially a freely falling body over the time period of integration.  We thus focus on frequencies satisfying
\be
\omega_m \ll \nu < \kappa.
\ee
In this regime, assuming that the damping of the mechanical resonator is small ($\gamma \ll \omega_m$), we can approximate the noise PSD as
\begin{equation}
\begin{split}
\label{relevantnoise}
N(\nu) &\approx \tilde{N}_{BM} + \Theta \nu^2.
\end{split}
\end{equation}
Here
\begin{equation}
\begin{split}
\tilde{N}_{BM}
&= N_{BM} +  \frac{4 \hbar^2    |G |^2 L  }{\kappa} \\
    \Theta
&= \hbar m \left[\frac{m \kappa^3}{4 \hbar |G|^2  (4+\kappa t_d)^2 } + \frac{4 \hbar   |G |^2  (4+\kappa t_d)^2}{m \kappa^3}  \right].
\end{split}
\end{equation}
We see that there is a white noise contribution (i.e.\ a renormalization of the thermal noise) as well as frequency-dependent contributions from both the shot and back-action terms. Above $\nu \gtrsim \sqrt{\kappa/2t_d}$, the shot noise term begins to dominate. At these high frequencies, back-action evasion is not effective, and the shot noise term is dominated by the cavity and mechanical response functions.

For small loss $L \approx 0$, we can minimize the noise \eqref{relevantnoise} with respect to the optomechanical coupling strength $G$, i.e.\ by varying the laser power $P$. Differentiating, one finds the optimized coupling
 \be
 \label{optimizedG}
 |G_{\rm opt}|^2 \approx \frac{1}{4}\frac{m \kappa^3}{\hbar (4+\kappa t_d)^2}.
\ee
Using this, the noise PSD \eqref{relevantnoise} becomes
 \begin{equation}
 \label{optimizednoise}
     N(\nu) \approx N_{BM}+\frac{\hbar m}{t_d^2} \left(  L + \nu^2 t_d^2 \right).
 \end{equation}

We can compare our results to those of the speedmeters being developed for gravitational wave detection, e.g. \cite{danilishin2018new}. Since the force and position noise spectral densities are related by a simple transfer function, $S_{xx} = |\chi_m|^2 S_{FF}$, it is clear that the back-action evasion our protocol achieves is proportionally the same whether reported in position or force units. We obtain a similar basic pattern of noise reduction as \cite{danilishin2018new}: sub-SQL noise in a reasonable band of frequencies, which must be targeted to the desired signal. For gravitational wave detection, the frequencies of interest are near the audio band; here, we are focused on much higher-frequency (radio band) signals, so we have chosen our system parameters accordingly. Moreover, we are focusing here on substantially smaller devices--milligram scale, compared to the gravitational speedmeter experiments with up to 200 kg mirrors, and accordingly smaller cavity lengths. In practice this is a much easier regime for dealing with issues of loss. The basic ideas here could potentially be demonstrated in a chip-scale device as proof of principle before being scaled to macroscopic devices.

\subsection{Impulse inference}

Our goal is to measure the net impulse delivered to the detector. We interferometrically read out the output light phase $Y'_{out}(t)$ as a time series, and then process this data to infer the impulse. The simplest option would be to consider the observable
\begin{equation}
\label{signal}
I(\tau) = \int_{0}^{\tau} F(t)dt,
\end{equation}
where $\tau$ is some integration time we can choose, and $F(t)$ is estimated from the observed $Y'_{out}(t)$ using \eqref{FE}. The noise in this signal is characterized by the variance,
\be
\label{rmsimpulse}
    \braket{\Delta I^2(\tau)} =
    \int_{-\infty}^{\infty} d\nu \frac{4\sin^2{(\nu \tau/2)}}{\nu^2}N(\nu).
\ee
This equation says that the RMS net impulse $\Delta I(\tau)$ delivered to the device purely by noise is calculable from the force noise PSD. As we have seen above, $N(\nu)$ has power law divergences at both high and low frequencies. In \eqref{rmsimpulse}, the sinc function will provide a cutoff on the high-frequency divergence, but the low-frequency divergence is still present. This means that we would naively infer that an infinite random impulse was delivered to the device! Of course, the actual physical impulse is finite. The divergence comes from the shot noise, i.e.\ counting statistics in our readout photons, and represents the fact that at low frequency this noise becomes arbitrarily large.

This suggests that we use a more intelligent observable than simply the integrated force. Suppose that we are looking for signals of a known shape in time $F_{\rm sig}(t)$. We want to test for the presence of this signal in our data $F(t)$. To do this we construct an observable $O(t)$ by filtering our data, where the filter scans over different possible event times $t_e$:
\begin{equation}
    O(t_e) = \int f(t_e -t^\prime)F(t^\prime) dt^\prime.
\end{equation}
By the convolution theorem, for a fixed time (without loss of generality, we can assume $t_e =0$), the variance in this estimator is,
\be
  \braket{\Delta O^2}  = \int |f(\nu)|^2 N(\nu) d\nu.
\ee
For example, a box filter $f(t) = \Theta(t-\tau) - \Theta(t)$ reproduces the simple estimator \eqref{rmsimpulse}. The signal to noise ratio (SNR) is then defined by
\begin{equation}
    \text{SNR}^2 = \frac{|\int f^*(\nu) F_{\rm sig}(\nu) d\nu|^2}{\int |f(\nu)|^2 N(\nu) d\nu}.
\end{equation}
Now we need a specific filter function which optimizes this signal to noise ratio. One can show (see appendix \ref{appendix-signal}) that the SNR is optimized by the filter 
\be
\label{filteropt}
f_{\text{opt}}(\nu) = \frac{F_{\rm sig}(\nu)}{N(\nu)}.
\ee
This is sometimes referred to as ``template matching'' \cite{braginsky1995quantum, brunelli1997template}. It says that the optimal filtering protocol is to scan for the expected signal shape renormalized by the noise model. With this choice of filter, the signal to noise ratio is given simply by
\be
\label{SNR}
\text{SNR}^2_{\text{opt}} =\int_{0}^{\infty} \frac{|F_{\rm sig}(\nu)|^2}{N(\nu)} d\nu.
\ee
Here we see a more robust interpretation of the divergences in our noise PSD: the very low and very high frequency parts of the spectrum make no contribution to the SNR.

\section{Example signal-to-noise calculations}

With our measurement protocol and corresponding noise PSD, we can now study our ability to detect particular signals. We begin with an instantaneous force, and then move on to the case of momentum transfer into a sensor by a passing object coupled to the sensor through a long-range force, for example gravity.

\subsection{Instantaneous force}

Consider an instantaneous force signal
\begin{equation}
    F_{\text{sig}}(t)= \Delta p \delta(t-t_0).
\end{equation}
This is a flat function of frequency, $F_{\rm sig}(\nu) = \Delta p e^{i \nu t_0} /\sqrt{2\pi}$. To estimate the signal-to-noise ratio achievable for such a signal, consider the optimized filter result \eqref{SNR}. As discussed above, our noise PSD starts to diverge for $\nu \gtrsim \sqrt{\kappa/2t_d}$. There is also a low-frequency divergence starting around the mechanical frequency $\nu \approx \omega_m$. In practice, there can also be some low-frequency cutoff set by a maximum integration time; for example, if we are trying to resolve individual impacts to a sensor which occur at some rate $\tau_{\rm coll}$, the signal from an individual event can only be obtained from frequencies $\nu \gtrsim 1/\tau_{\rm coll}$. Given these limits, we can then approximate (in fact, lower-bound) the SNR, using \eqref{SNR} and \eqref{relevantnoise}, as
\begin{align}
\begin{split}
\label{SNRinstant}
{\rm SNR}^2 & \gtrsim \frac{\Delta p^2}{2\pi} \int_{\tau_{\rm coll}^{-1}}^{\sqrt{\kappa/2t_d}} \frac{d\nu}{\tilde{N}_{BM}+\Theta \nu^2} \\
& \approx \frac{\Delta p^2}{4\sqrt{\Theta \tilde{N}_{BM}}}.
\end{split}
\end{align}
The approximation in the last line holds if $\sqrt{\Theta \kappa/2 \tilde{N}_{BM} t_d} \gg 1 \gg \sqrt{\Theta/\tilde{N}_{BM} \tau^2_{\rm coll}}$. The first condition says that the detector can move information between cavities fast, and the second says that the collisions do not occur too rapidly, in comparison to the typical time scales occurring in the noise.

As an example, consider the following problem: can we use this protocol to count individual gas collisions with a sensor? Concretely, imagine that we place our sensor in a vacuum chamber and continuously monitor it with our protocol. Let's assume that the sensor is freely-falling, or at least that the mechanical damping $\gamma$ is so low that we can ignore phononic loss into the support, so that the only ``thermal noise'' comes from individual gas collisions with the device. We would then view the noise PSD as coming strictly from the quantum measurement-added noise, and the gas collisions are actually the signal we try to detect above the noise. We thus have $N(\nu) = \hbar m (L/t_d^2 + \nu^2)$, and the SNR for a single gas collision transferring a momentum $\Delta p$ (taken to occur instantaneously) is then just the $\gamma \to 0$ limit of \eqref{SNRinstant},
\be
\label{SNRgas}
{\rm SNR}^2 = \frac{\Delta p^2 t_d}{\sqrt{L} \hbar m }.
\ee
This simple answer has a satisfying interpretation: the factor $\hbar m/t_d$ is what one would naively obtain for a standard quantum limit on impulse sensing over a time $t_d$ (cf. the expression for SQL position uncertainty $\Delta x_{SQL}^2 = \hbar t/m$). We are then seeing a noise below the SQL by a factor of the delay line loss $L^{1/4}$, the limiting factor in our protocol. This represents the central idea in this paper: inferring force through a direct measurement of the momentum out-performs use of a position measurement. In the next section, we will see the same behavior for a different problem, the detection of impulses from long-ranged forces.

Numerically, we have in this limit
\begin{align}
\label{SNR-gas-estimate}
{\rm SNR} \approx 1 \times \left(\frac{\Delta p}{10 ~ \rm{keV/c}}\right) \left(\frac{1~\rm{fg}}{m}\right)^{1/2} \left(\frac{10^{-4}}{L}\right)^{1/4},
\end{align}
assuming a delay time $t_d \sim 10^{-5} ~{\rm sec}$.\footnote{This would require an extremely long optical fiber, but could also be achieved for example by using a third cavity as the ``delay line''.} Consider helium gas at room temperature. If the atoms scatter elastically off the sensor, the typical momentum transfer should be of the order $\Delta p \sim \sqrt{m_{\rm gas} k_B T} \sim 10~{\rm keV/c}$. Thus, with a femtogram-scale detector (e.g.  \cite{lahaye2004approaching,sun2012femtogram,de2017universal}), we would have the ability to resolve the individual gas collisions above the measurement noise. 

Qualitatively, this calculation illuminates a fundamental limitation to momentum sensing. When we look for our signal, we assume some kind of template fitting, as discussed in the previous section. Naively, one might have expected that the best strategy to detect an instantaneous force would be to use a template that is essentially itself a delta function in time. But this is not right: if one only integrates the signal instantaneously, the detection will be limited by shot noise, and in fact the SNR is strictly zero. Quantitatively, we can see that our optimal filter \eqref{filteropt} has a bandwidth $\Delta \nu \approx \sqrt{\kappa/2t_d}$, and thus finite support as a function of time.

\subsection{Long range forces}
\label{longrange}

Now we consider detection of some object approaching the sensor and interacting with it through a long range force. Our primary motivation here is the gravitational detection of passing dark matter \cite{Carney:2019pza}, but the problem can be phrased more generally. We consider a $1/r$ potential between sensor and incoming particle 
\begin{equation}
\label{long_pot}
    V(t)=\frac{\beta}{r(t)}.
\end{equation}
Here $\beta$ is a constant with dimensions of ${\rm energy} \times {\rm length}$ which characterizes the long range force. For example, $\beta= Q_1 Q_2/4\pi$ for the Coulomb force between two charges, or $\beta = G_N m_1 m_2$ for the Newtonian gravitational force.

\begin{figure}[t]
    \centering
    \includegraphics{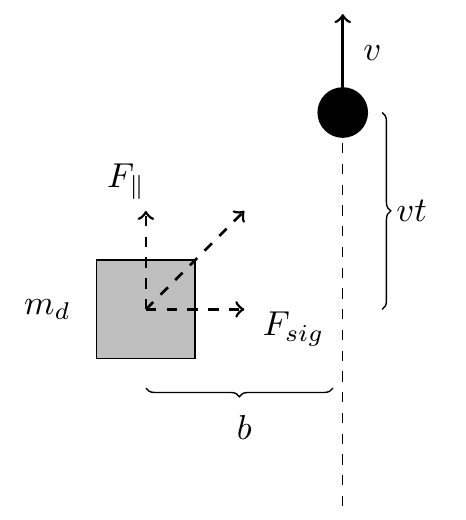}
    \caption{Kinematics of the long-range scattering event. A particle passes near the sensor, with impact parameter $b$ and velocity $v$. This leads to an effective interaction time $\tau \sim b/v$.}
    \label{figure-scattering}
\end{figure}

\begin{figure}[t]
\includegraphics[scale=.5]{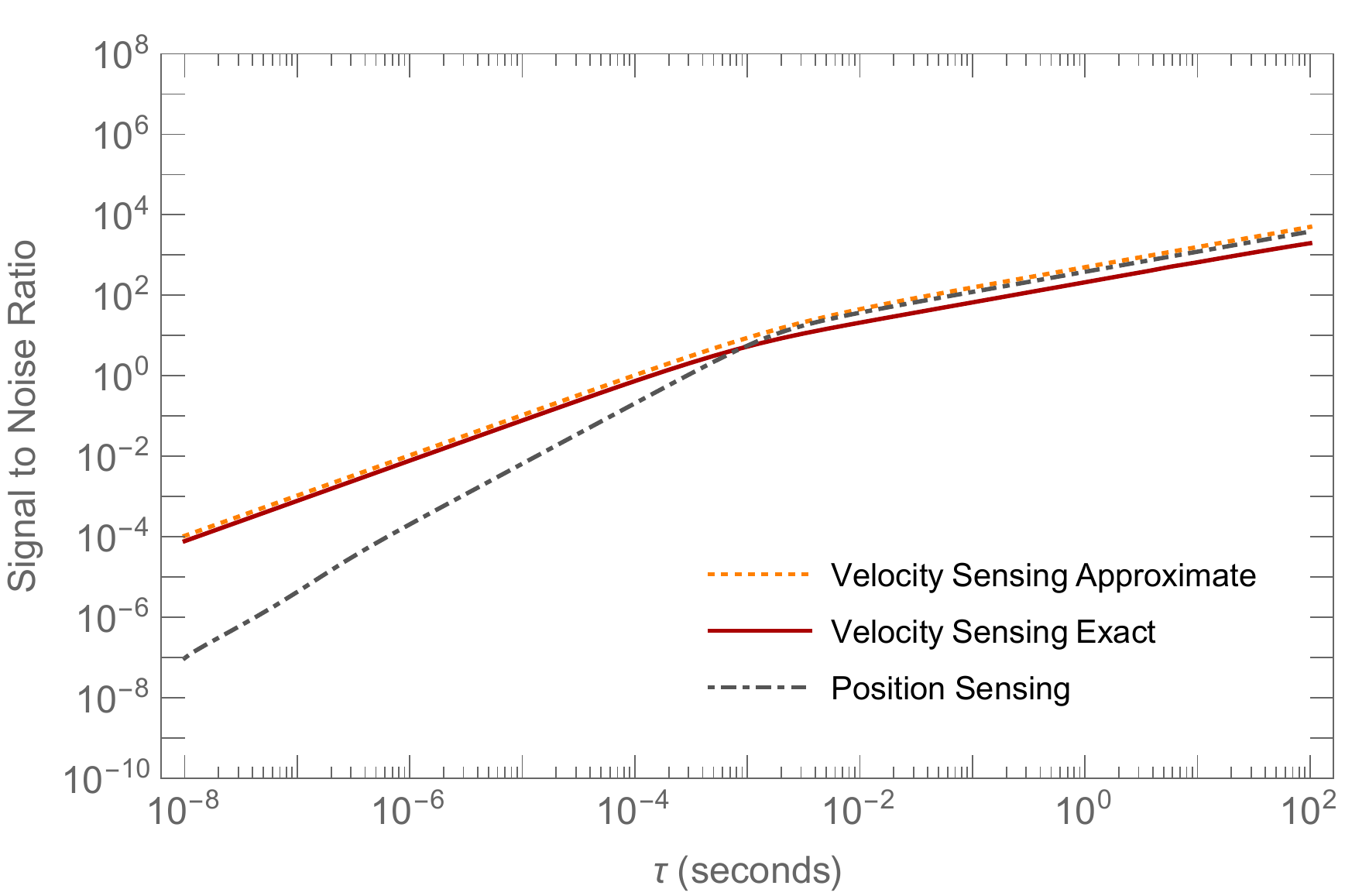}
\caption{Signal to noise ratio for long range force detection, as a function of fly-by time $\tau$. We see a clear improvement in the SNR obtained with our velocity sensing protocol \eqref{noisepsd} compared to the result with position sensing \eqref{noisepsdsingle}, for fast signals. For slow signals (here, $\tau \gtrsim 1~{\rm ms}$), the measurement-added noise becomes subdominant to the thermal noise, and our back-action-evasion scheme does not help. To check our approximate result \eqref{approxSNR}, we also display the exact SNR calculated using the full noise PSD \eqref{noisepsd} and signal \eqref{exactsig}. Here we use the same detector parameters as in figure \ref{figure-noisecontribs} and at optimized G for both cases [see \eqref{optimizedG},\eqref{optimizedGsingle}], and use the gravitational dark matter signal as in  \eqref{SNR-longrange} with a very heavy dark matter candidate $m_\chi = 10 ~\rm{mg}$ and impact parameter $b = 1 ~\rm{mm}$.}
\label{figure-SNR}
\end{figure}

Consider a particle passing by the detector at a high velocity $v$ and impact parameter $b$ interacting with a sensor via \eqref{long_pot}, as in figure \ref{figure-scattering}. For simplicity, we assume that the particle's trajectory is a straight line. Almost all of the momentum is transferred to the sensor over a timescale $\tau \sim b/v$ when the particle is nearest to the sensor. The force acting on the sensor has pieces parallel and perpendicular to the particle track. The parallel component of the force exerted over this time period transfers no net momentum to the sensor.\footnote{We note however that one could in principle use an antisymmetric-in-time filter to look for this signal.} Thus we focus on the perpendicular component of the force,
\begin{equation}
F_{\rm{sig}} = \frac{\beta b}{(b^2+v^2 t^2)^{3/2}}.
\end{equation}
 The Fourier transform of this signal is
\begin{equation}
\label{exactsig}
    F_{\rm{sig}}(\nu) = \sqrt{\frac{2}{\pi}}\frac{\beta |\nu|}{v^2} K_1\left(\frac{b}{v}|\nu|\right)
\end{equation}
where $K_1$ is a modified Bessel function. For the purposes of estimating a signal-to-noise ratio in \eqref{SNR}, we can approximate (in fact, underestimate) this signal using an exponential
\begin{equation}
F_{\rm{sig}}^{\rm{approx}}(\nu) = \sqrt{\frac{2}{\pi}} \frac{\beta }{b v} e^{(-\tau |\nu|/2)}.
\end{equation}

As discussed in the previous section, the noise PSD starts to diverge at frequencies above $\nu \gtrsim \sqrt{\kappa/2t_d}$. We can therefore approximate the signal-to-noise ratio \eqref{SNR} here as
\begin{equation}
\label{approxSNR}
\begin{split}
\rm{SNR}^2 & \approx  \frac{2 \beta^2}{\pi b^2 v^2} \int^{\sqrt{\kappa/2 t_d}}_0 \frac{e^{-\tau \nu}}{\tilde{N}_{BM}+ \Theta \nu^2} d\nu \\
& \gtrsim \frac{ \beta^2}{ b^2 v^2 \tau \tilde{N}_{BM}} \frac{1}{(1+\eta)}.
\end{split}
\end{equation}
The approximation in the second line is good as long as  $\tau \sqrt{\kappa/2t_d} \gtrsim 1$. The dimensionless parameter $\eta$ is defined as
\be
\eta = \sqrt{\Theta/\tilde{N}_{BM} \tau^2} \gg 1,
\ee
with this inequality holding in examples, to which we now turn. We have numerically verified that these approximations to the exact SNR calculated using the full noise PSD \eqref{noisepsd} and signal \eqref{exactsig} are highly accurate, see figure \ref{figure-SNR}.

We now use this formalism to show the central result of this paper: momentum monitoring out performs position monitoring for measuring rapid impulses. Specifically,
with $G$ optimized as in \eqref{optimizedG} and with small enough loss coefficient so that $\tilde{N}_{BM} \approx N_{BM} = 4  \gamma m k_B T$, we can write the parameter $\eta^2 = (\hbar m/\tau)/(N_{BM} \tau)$. If our measurement protocol had SQL-level measurement noise (see appendix \ref{appendix-singlesided} for details), the variance in our measured impulse would be given by
\be
\Delta p_{\rm noise}^2 = N_{BM} \tau + \frac{\hbar m}{\tau} = N_{BM} \tau (1 + \eta^2).
\ee
Here the first term comes from thermal noise and the second from the SQL measurement noise. In contrast, reading off the denominator of \eqref{approxSNR}, we find with our protocol that the noise is
\be
\label{deltap2good}
\Delta p_{\rm noise}^2 = N_{BM} \tau (1 + \eta),
\ee
thus we have a reduction in noise by a factor $1/(1+\eta) \ll 1$. This is the analogue in the long-range detection problem to the noise reduction displayed in \eqref{SNRgas}.

This result could be further improved by including the benefits of using squeezed light for detection. The noise above arises from assuming that the $X$ and $Y$ quadratures noises are uncorrelated, which is an assumption that is broken for squeezed light. For position monitoring, the appropriate quadrature and amplitude of light -- necessary choices for using squeezing to improve the measurement -- depend sensitively on the bandwidth and target frequency of the signal. That is, shot noise and back-action in those settings scale differently with frequency. Here, however, the frequency dependence from $\omega_m$ up to $\sim \kappa$ is the same for both quadratures, and thus broadband squeezed light suffices.
 
As a numerical example, we now consider explicitly the gravitational detection of a passing particle, for example a heavy dark matter candidate \cite{Carney:2019pza}. The key scaling property in this problem is that the signal strength scales linearly in both the sensor mass $m_s$ and dark matter mass $m_{\chi}$, and is enhanced by small impact parameters as $1/b^2$. In contrast, the noise scales like $\sqrt{m_s}$. In a terrestrial experiment, individual dark matter particles pass through the lab at the ``wind speed'' $v_{DM} \sim 220~{\rm km/s}$. Considering a fiducial impact parameter on the millimeter scale then leads to a very short flyby time $\tau \sim 10^{-8}~{\rm sec}$. In this setting we obtain an SNR 
\begin{equation}
\label{SNR-longrange}
\begin{split}
\rm{SNR}  & \approx \frac{ G_N m_{\chi} m_s}{ b v \sqrt{\tau N_{BM}}} \frac{1}{\sqrt{(1+\eta)}}\\
& \approx 10^{-3} \times \left(\frac{m_\chi}{10 ~ \rm{mg}}\right) \left(\frac{m_s}{1~\rm{ g}}\right)^{\frac{1}{2}} \left(\frac{\tau}{10 ~\rm{ns}}\right)\left(\frac{1 ~\rm{mm}}{b}\right)^2,
\end{split}
\end{equation}
where we take the thermal Brownian noise at dilution refrigeration temperature $T \sim 10~{\rm mK}$ and assumed a very high-$Q$, low-frequency resonant detector with $\gamma \sim 10^{-4} ~{\rm Hz}$. The scaling with $\tau$ in the numerical estimate here is valid for signals fast enough that $\eta \gsim 1$, in which case the SQL level measurement added noise is greater than the thermal noise in the system $\hbar m/\tau^2 \ge N_{BM}$. In this example, the crossover occurs around $\tau \sim 1~{\rm ms}$, as can be seen from figure \ref{figure-SNR}.

From this estimate, we see that this measurement protocol is not yet sensitive enough for gravitational detection of Planck scale ($m_{\chi} \sim m_{P} \sim 10 \ {\rm \mu g}$) dark matter particles. A more sophisticated protocol will be necessary to achieve the goals outlined in \cite{Carney:2019pza}, for example, using squeezed light. However, a device with the sensitivity given here could be used, for example, to exclude dark matter models which couple through some other long range force a few orders of magnitude stronger than gravity, for example the modified gravity models in \cite{hall2018laser} or some composite dark sector models coupled through a new light gauge boson (e.g. \cite{Krnjaic:2014xza,Gresham:2017zqi,coskuner2019direct}). Bounds on these types of dark matter models coming from impulse sensing detectors will be studied in detail in a future publication.

\section{Conclusions and Discussions}

Detection of a rapidly delivered impulse is a ubiquitous problem in many branches of physics. Fundamental quantum measurement noise in such a detection is often the ultimate limitation to reaching better sensitivities. In this paper, we have demonstrated that the use of a direct momentum sensing protocol can significantly reduce this noise in comparison with the more traditional approach of position sensing.

Here, we have presented a concrete example of this general phenomenon using an optomechanical system involving a pair of cavities probed continuously by a laser. This specific approach is ultimately limited by optical losses in a delay line which transmits the probe light between the cavities. With currently available fabrication techniques, these losses limit the noise reduction in this protocol to around 30 dB below the standard quantum limit. Other protocols, for example involving discrete pulse sequences \cite{khosla2017quantum} or direct measurement of velocity through an inductive coupling \cite{wagoner1979tunable} could improve the situation, and require more detailed future study.

Given that momentum measurement or impulse detection is commonly needed, the results presented here could have wide applications. We gave a pair of examples, one in metrology and the other in particle physics. In the former, we suggested that our protocol is already sensitive enough [see equation \eqref{SNR-gas-estimate}] to monitor all of the individual gas particles colliding with a femtogram-scale sensor in a room temperature, high vacuum environment. This could be used for example in quantum-limited pressure calibrations. In the latter, we studied the application of this protocol to the detection of heavy dark matter candidates purely through their gravitational interaction with a sensor \cite{Carney:2019pza}. Although the sensitivity of the simple protocol presented here [see equation \eqref{SNR-longrange}] is too limited by optical losses to achieve the requirements of \cite{Carney:2019pza}, this study shows a clear path to straightforward improvements, which we leave to future work. We hope that the example studied in this paper serves to guide the way to impulse measurement schemes reaching the fundamental limits allowed by quantum mechanics, enabling detection of such extremely weak signals.

\section*{Acknowledgements}

We thank Nancy Aggarwal, Aashish Clerk, Jonathan Kunjummen, Nergis Mavalvala, and Kartik Srinivasan for helpful conversations. DC thanks the Les Houches school ``Quantum Information Machines'' and the Galileo Galilei Institute  workshop ``Next Frontiers in the Search for Dark Matter'' for hospitality while a portion of this work was completed. Fermilab is operated by Fermi Research Alliance, LLC, under Contract No. DEAC02-07CH11359 with the US Department of Energy. SG is supported by the Physics Frontier Center at the Joint Quantum Institute, which is funded through the National Science Foundation (Award no. 1430094).

\medskip

\appendix

\section{Continuous position measurement in a single sided cavity}
\label{appendix-singlesided}

\begin{figure}[t]
\includegraphics[scale=1]{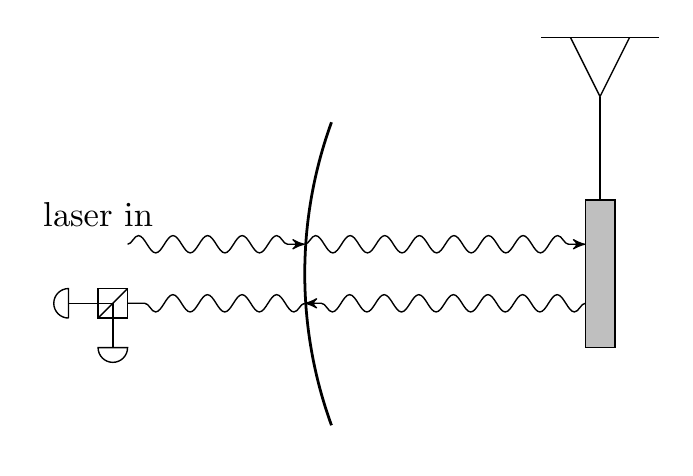}
\caption{Schematic of a single-sided cavity optomechanics experiment. The cavity is driven by a laser from outside. The light picks up a phase $\phi \propto x(t)$ proportional to the mechanical position, which is then read out through an interferometer after the light exits the cavity.}
\label{single-sided}
\end{figure}

In this appendix, we present a detailed formalism for detection of forces using a prototypical single-sided optomechanical system implementing a continuous position measurement. Our treatment borrows heavily from  \cite{RevModPhys.82.1155}, and we refer the reader to that review for further details.

The optomechanical system consists of a partially transparent fixed mirror on one side and another movable/suspended perfect mirror on the other side as in figure \ref{single-sided}, forming a cavity whose frequency is a function of the position $x = x(t)$ of the movable mirror. The cavity mode, mirror, and their respective baths can be characterized by the total Hamiltonian
\begin{equation}
H_{\text{tot}} = H_{\text{cav}}+H_{\text{mech}}+H_{\text{bath}}.
\end{equation} 
Both the cavity mode and mirror are modeled as harmonic oscillators,
\begin{align}
\begin{split}
H_{\text{cav}}&= \hbar \omega (x) a^\dagger a\\
H_{\text{mech}}&= \frac{1}{2}m \omega_m^2 x^2+\frac{p^2}{2m}.
\end{split}
\end{align}
Here $a$ denotes the annihilation operator for the optical cavity mode, and $\omega$ and $\omega_m$ are the resonance frequencies of the cavity and resonator respectively. The cavity resonance frequency $\omega = \omega(x)$ is a function of the length of the cavity and it changes as the mechanical resonator at one end moves. This interaction couples the cavity and the resonator. For small displacements of the mirror, we can Taylor expand the position-dependent cavity frequency
\be
\omega(x) = \omega_c \left( 1 - \frac{x}{\ell}  + \Oi(x^2) \right),
\ee
where $\omega_c = 2\pi c/\ell$ is the cavity frequency when the mirror is at rest, $\ell$ being the equilibrium length of the cavity. Using this result, the optomechanical interaction can then be characterized with the following Hamiltonian:
\begin{equation}
    H_{\rm{int}}= \hbar g_0 \frac{x}{x_0} a^\dagger a.
\end{equation}
 The interaction strength here is defined as $g_0 =x_0 \frac{d\omega}{dx} = -x_0 \omega_c/\ell$. Here $x_0$ is an arbitrary length scale which we factor out so that $g_0$ has units of a frequency. This interaction couples the cavity photons to the mechanical position and is the key to prepare and read out the mechanical motion through the output light.
 
Both the cavity and the mechanical system have their own baths. The cavity bath consists of photons which are outside the cavity and can enter and exit through the fixed mirror. The mechanical bath includes degrees of freedom like ambient gas particles which can collide with the mechanics or phonons in the support structure suspending the movable mirror. Both baths consist of a large number of modes. In general, we write the bath Hamiltonian and coupling to the system as 
\begin{equation}
\begin{split}
H_{\rm{bath}} &= \Sigma_p \hbar \omega_p A_p^\dagger A_p + \Sigma_p \hbar \nu_p B_p^\dagger B_p \\
&   - i\hbar \Sigma_p \left[f_p a^\dagger A_p -  f_p^\star a A_p^\dagger\right] - i\hbar\Sigma_p \left[g_p b^\dagger B_p -g_p^\star b B_p^\dagger\right].
\end{split}
\end{equation}
Here the $A_p, B_p$ are the cavity and mechanical bath modes, indexed by an arbitrary label $p$, and $\omega_p,\nu_p$ are the frequencies of these modes. We have have also introduced the mechanical annihilation operator $b$, and the coupling constants $f_p,g_p$.

The bath modes can be integrated out by solving their equations of motion explicitly in terms of their initial conditions and the system variables. Within the Markovian approximation, we can define the bath ``input operators'' \cite{gardiner1985input,RevModPhys.82.1155}
\begin{equation}
\begin{split}
A_{\text{in}}(t)&= \frac{1}{\sqrt{2 \pi \rho_A}}\Sigma_p e^{-i \omega_p (t-t_0) A_p (t_0)}\\
B_{\text{in}}(t)&= \frac{1}{\sqrt{2 \pi \rho_B}}\Sigma_p e^{-i \nu_p (t-t_0) B_p (t_0)},
\end{split}
\end{equation}
where $\rho_A, \rho_B$ are the densities of states of the baths. Assuming the couplings are constant for the modes of interest $f_p \equiv f, g_p \equiv g$, these quantities are related to the cavity and mechanical energy loss rates via $\kappa = 2 \pi f^2 \rho_A, \gamma=2 \pi g^2 \rho_B$ respectively. We then define the input  mechanical force $F_{\rm in}$ in terms of  $B_{in}$ and  $B_{in}^{\dagger}$. The input force consists of the deterministic signal $F_{\rm sig}(t)$ plus random Brownian noise, which we model as usual thermal (Johnson-Nyquist) white noise \eqref{thermalnoise}.

The cavity mode, on the other hand, will be driven by an external laser. In other words, we take the cavity input modes  to consist of fluctuations around a classical background. This effectively displaces the cavity operators by $a\rightarrow (\alpha+a )e^{-i \omega_L t}$, with $\omega_L$ being the frequency of the monochromatic laser and $\alpha \propto \sqrt{P/\hbar \omega_L \kappa}$ being the drive strength in terms of the laser power $P$ and cavity energy loss rate $\kappa$. We move to a frame co-rotating with the drive by applying a unitary transform  $U = e^{i \omega_L a^\dagger a t}$ to the Hamiltonian. This modifies the cavity Hamiltonian to
\begin{equation}
H_{\text{cav}} = -\hbar \Delta a^\dagger a 
\end{equation}
where $\Delta = \omega_L-\omega_c$, is the detuning due to the drive. Here we will work on resonance when $\Delta=0$. For a strong drive, we can linearize the Hamiltonian in the fluctuations,
\begin{equation}
H_{\rm{int}} = \hbar g_0 \alpha \frac{x}{x_0} (a + a^\dagger) + \hbar g_0  \alpha^2 \frac{x}{x_0}.
\end{equation}
The second term here is just a constant radiation pressure which shifts the equilibrium position of the mechanical resonator. We can re-absorb this into the definition of the constants, thus we drop this term in the following. Defining the quadratures of the cavity to be $ \hat{X}= (\hat{a}+\hat{a}^\dagger)/\sqrt{2}$ and $ \hat{Y}= -i(\hat{a}-\hat{a}^\dagger)/\sqrt{2}$, we have the commutation relation $[\hat{X},\hat{Y}] = i$, and obtain the effective optomechanical interaction Hamiltonian, 
\begin{equation}
 H_{\rm{int}} =  \hbar G x X
\end{equation}
where the effective optomechanical coupling strength is defined as $G = \sqrt{2} g_0 \alpha/x_0$ which has the dimension of frequency per length. Here we have chosen a gauge where the coupling is purely between the mechanical position $x$ and optical amplitude quadrature $X$ for notational simplicity.

All told, we can now write down the Heisenberg-Langevin equations of motion for the optical and mechanical quadratures. These read
  \begin{align}
\begin{split}
\dot{X} &= -\frac{\kappa}{2}X+ \sqrt{\kappa} X_{in}\\
\dot{Y} &= -G x -\frac{\kappa}{2}Y+ \sqrt{\kappa} Y_{in}\\
\dot{p} &= -\hbar G X - \gamma p +  F_{in}-m\omega_m^2 x\\ 
\dot{x} &= \frac{p}{m}.
\end{split}
\end{align}
Here, the input optical quadratures are defined as $X_{\rm in} = (A_{\rm in} + A_{\rm in}^{\dagger})/\sqrt{2}, Y_{\rm in} = -i (A_{\rm in} - A_{\rm in}^{\dagger})/\sqrt{2}$. These represent the vacuum fluctuations of the light around the classical laser drive, and are taken to satisfy white noise correlation functions of the form
\begin{align}
\begin{split}
\label{vacuum-single}
\braket{X_{in}(t) X_{in}(t')} & = \braket{Y_{in}(t) Y_{in}(t')} =  \delta(t-t') \\
\braket{X_{in}(t) Y_{in}(t')} & = 0.
\end{split}
\end{align}
Note that this assumes the fluctuations are not correlated; these relations are modified in the presence of non-trivial input states, for example squeezed light.

Each input field has a corresponding output field. These are related by the input-output relations
\begin{align}
\begin{split}
\label{input-output-single}
X_{\text{out}} &= X_{\text{in}} - \sqrt{\kappa} X\\
Y_{\text{out}} &= Y_{\text{in}} - \sqrt{\kappa} Y.
\end{split}
\end{align}
The output phase quadrature $Y_{\rm out}$ is what we have experimental access to, via an external homodyne interferometer. Thus, we want to solve for $Y_{\rm out}$ in terms of the various input fields. This is trivial in the frequency domain since the equations of motion are linear. In terms of the mechanical and cavity response functions \eqref{responsefns}, one finds easily that
 \be
 \label{Youtsimple}
Y_{\text{out}} = e^{i \phi_c} Y_{\text{in}} + G  \chi_c \chi_m \left[F_{\text{in}} -\hbar G \chi_c X_{\text{in}} \right].
\ee
Here we defined the ``cavity phase shift''
\be
e^{i \phi_c} = 1-\sqrt{\kappa} \chi_c = \frac{-i \nu - \kappa/2}{-i \nu + \kappa/2}.
\ee
To estimate the force $F$ on the mechanics given our observed $Y_{\rm out}$, we can simply divide through with the appropriate coefficient and define an estimator for the force:
\begin{equation}
\label{FE-single}
F_E = \frac{ Y_{out} }{G   \chi_c \chi_m }.
\end{equation}

As discussed in the main text, the noise in the measurement protocol is characterized by the noise power spectral density (PSD), defined as in \eqref{FEPSD}. Here, using \eqref{Youtsimple} and the various noise correlation functions, we obtain
\begin{equation}
\label{noisepsdsingle}
\begin{split}
N(\nu) &=\frac{1 }{  |G|^2 |\chi_c|^2 |\chi_m|^2 } +  N_{BM} +  \hbar^2  |G|^2   |\chi_c|^2.
\end{split}
\end{equation}
The first term here is the shot noise arising from the statistical counting errors of photons. The middle term is the thermal noise. The last term denotes the back-action noise arising from the light randomly pushing the mirror around while probing the system.
 
Here, we notice that the shot noise term is inversely proportional to $G$ whereas the back-action term is directly proportional to $G$. We can thus optimize these two terms with respect to the optomechanical coupling strength $G$ which is in turn dependent on the laser power. Note that this procedure is done at some particular fixed frequency. Differentiating with respect to $G$ one finds the optimum
\begin{equation}
\label{optimizedGsingle}
 |G_{\rm{opt}}|^2 = \frac{1}{\hbar   |\chi_c|^2 |\chi_m| }.
\end{equation}
For a sinusoidal signal of frequency $1/\tau$ and assuming a small damping coefficient $\gamma$ in the band of $\nu \gg \omega_m$, the noise spectrum is well approximated by
\begin{equation}
 N(1/\tau) \approx \frac{\hbar m}{ \tau^2} + N_{BM}.
\end{equation}
 The variance in the measured impulse with this measurement protocol would then be given by
\be
\Delta p_{\rm noise}^2 = N_{BM} \tau + \frac{\hbar m}{\tau} = N_{BM} \tau (1 + \eta^2)
\ee
where $\eta = \sqrt{\frac{\hbar m}{\tau^2 N_{BM}}}$. One can compare this expression to the one obtained in the momentum-measurement protocol, \eqref{deltap2good}. For the parameters of interest in this paper, we have $\eta \gg 1$, and we see that the momentum measurement protocol outperforms this SQL-level position measurement protocol by a factor of $\eta$.

\vspace{1.5pt}

\section{Signal processing and optimal filter theory} 
\label{appendix-signal}

Suppose we have some observed force signal $F(t)$ as a time series, which we have estimated from our output light. We want to test the hypothesis that there is a signal $F_{\rm{sig}}(t)$ of known shape in the data, occurring at some unknown event time $t_e$. We thus need to use a filter $f(t-t_e)$ (``template'') to scan through the data through convolution. We define our estimator for the signal
\begin{equation}
    O(t_e) = \int f(t_e -t)F(t) dt.
\end{equation}
The noise in this quantity is independent of the event time $t_e$. Taking $t_e = 0$ for simplicity and using the convolution theorem, we have
\begin{equation}
O(0) = \int f^*(\nu)F(\nu) d\nu.
\end{equation}
The variance is thus
\be
\braket{\Delta O^2}  &= \int |f(\nu)|^2 N(\nu) d\nu
\ee
using our definition of the noise PSD $N(\nu)$. Thus the signal to noise ratio is
\begin{equation}
    \text{SNR}^2 =  \frac{|\int f^*(\nu) F_{\rm{sig}}(\nu) d\nu|^2}{\int |f(\nu)|^2 N(\nu) d\nu}.
\end{equation}

The question is then: given a particular signal $F_{\rm sig}(\nu)$ and noise PSD $N(\nu)$, what is the optimal choice of filtering function $f(\nu)$ which maximizes the signal-to-noise ratio? Let us redefine the integration variable
\begin{equation}
    du = N(\nu) d\nu.
\end{equation}
Since $N(\nu) > 0$ is positive everywhere, this change of variable is a valid transformation. Using this we can rewrite the SNR as
\begin{equation}
    \text{SNR}^2 =  \frac{|\int f^*(u) F_{\rm{sig}}(u) \frac{d\nu}{du} du|^2}{\int |f(u)|^2 du}.
\end{equation}
We notice that we have an inner product of $f(u)$  with the function $F_{\rm{sig}}(u) d\nu/du$. We want to maximize this inner product while keeping the norm of $f$ fixed. This means that the functions are necessarily going to be parallel. Thus the optimal choice is
\be
f_{\rm opt}(u) = F_{\rm{sig}}(u) d\nu/du.
\ee
Inverting this back to frequency domain, we have the simple result for the optimal filter
\be
f_{\rm opt}(\nu) = \frac{F_{\rm{sig}}(\nu)}{N(\nu)}.
\ee
Using this filter, the signal-to-noise ratio is given by
\be
\text{SNR}_{\text{opt}} =\int \frac{|F_{\rm{sig}}(\nu)|^2}{N(\nu)} d\nu.
\ee

\section{Slight Detuning}

It is experimentally challenging to maintain both the ring cavities at the same equilibrium resonance frequency. We can quantify the impact of a slight imperfection by introducing a small amount of detuning to both the cavities. To get some intuition in a simple setting, we begin with a single sided cavity to understand how a slight detuning affects the quadratures of interest and the noise PSD. Then we move on to a qualitative discussion on the effects of detuning in our velocity measurement protocol with a double ring cavity, and provide some numerical estimates.

\subsection{Slight Detuning in a Single Sided Cavity}
With the introduction of detuning in the system, the single sided cavity Hamiltonian gets modified to
\begin{equation}
\label{Hamiltonian-single-detuned}
H= -\hbar \Delta a^\dagger a + H_{mech} + H_{int}.
\end{equation} 

Let us first consider this scenario in terms of the phase picked up by the cavity field. We are interested in the limit that the signal is much faster than both the mechanical period and the delay time. Thus we can consider the phase shifts picked up in the limit that the mechanical element is stationary.  For a detuned single-sided cavity,  the Hamiltonian is given in \eqref{Hamiltonian-single-detuned} where the interaction term is proportional to $\hbar g_0 a^\dagger a x/x_0$ before linearization. The quantum Langevin equation for the optical field in the cavity is
\begin{equation}
\dot{a}= i (\Delta -g_0 \frac{x}{x_0}) a -\frac{\kappa}{2} a + \sqrt{\kappa} a_{in}.
\end{equation}
We can define an effective detuning parameter
\be
\Delta_{\rm eff} = \Delta -g_0 \frac{x}{x_0}.
\ee
Then for a steady state solution in the cavity, we obtain the phase shift of the light
\begin{equation}
a= \frac{\sqrt{\kappa}}{- i \Delta_{\rm eff} +\kappa/2} a_{in}.
\end{equation}
Following the input-output relation for the optical field we obtain the output field
\begin{equation}
 \begin{split}
 a_{out}&=a_{in} -\sqrt{\kappa} a\\
 &= \frac{{-i \Delta_{\rm eff} -\kappa/2}}{{-i \Delta_{\rm eff} +\kappa/2}} ~ a_{in}  \\
 &= e^{i \phi} a_{in}.
 \end{split}
 \end{equation}
This phase shift formalism will be more important in our discussion of the double ring cavity in the next section. We can also use the familiar quadrature formalism to describe our detuned single sided  cavity. The equations of motion from the above Hamiltonian are
\begin{align}
\begin{split}
\label{EOMsingledetune}
\dot{X} &= -\frac{\kappa}{2}X+ \sqrt{\kappa} X_{in}- \Delta Y\\
\dot{Y} &= -G x -\frac{\kappa}{2}Y+ \sqrt{\kappa} Y_{in} + \Delta X\\
\dot{p} &= -\hbar G X  - m\omega_m^2 x -  \gamma p +  F_{in}\\ 
\dot{x} &= \frac{p}{m}.
\end{split}
\end{align}
Note that the optical quadratures are now coupled through this detuning parameter.

\begin{figure}[t]
\includegraphics[scale=0.45]{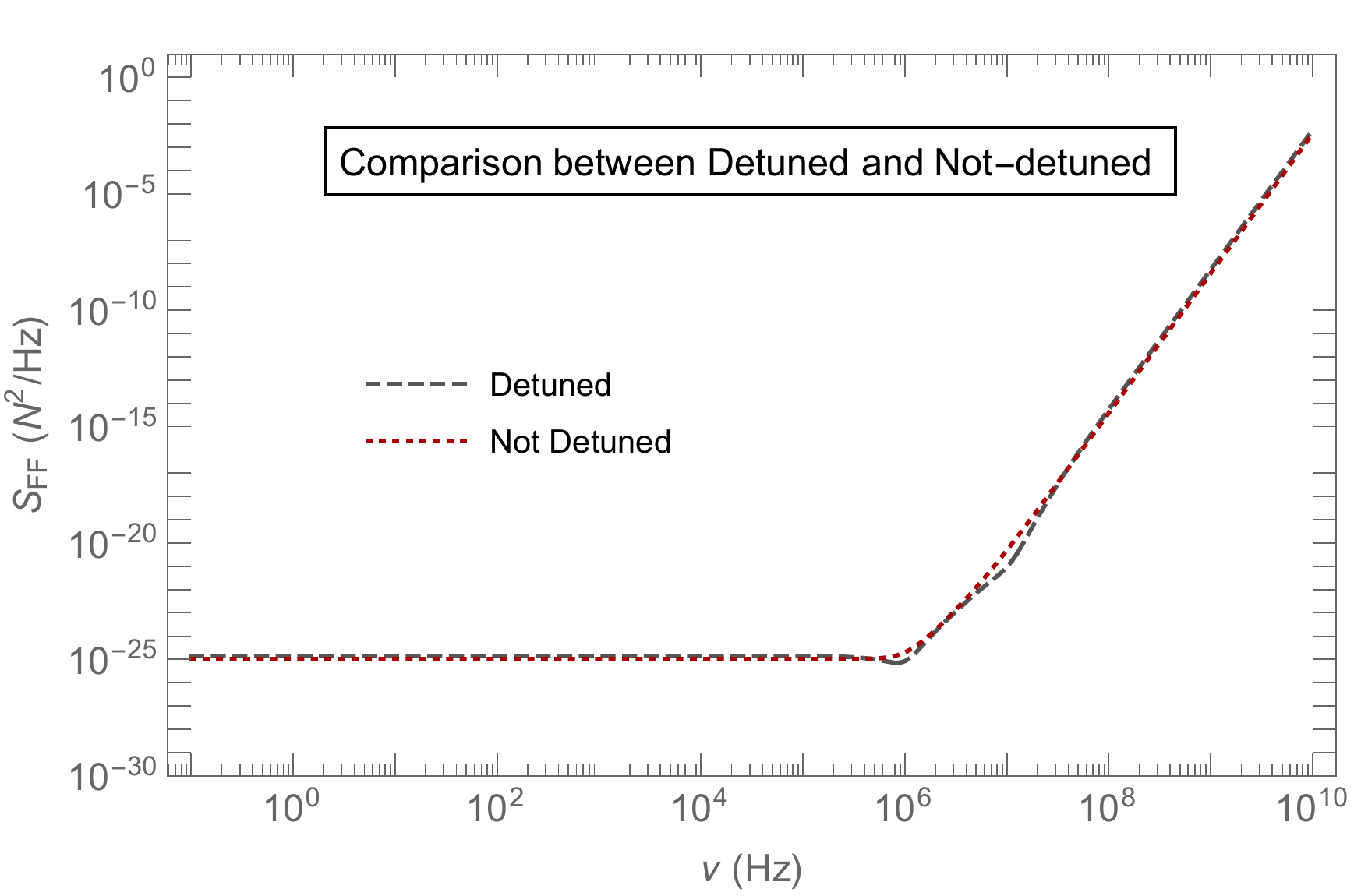}
\includegraphics[scale=0.43]{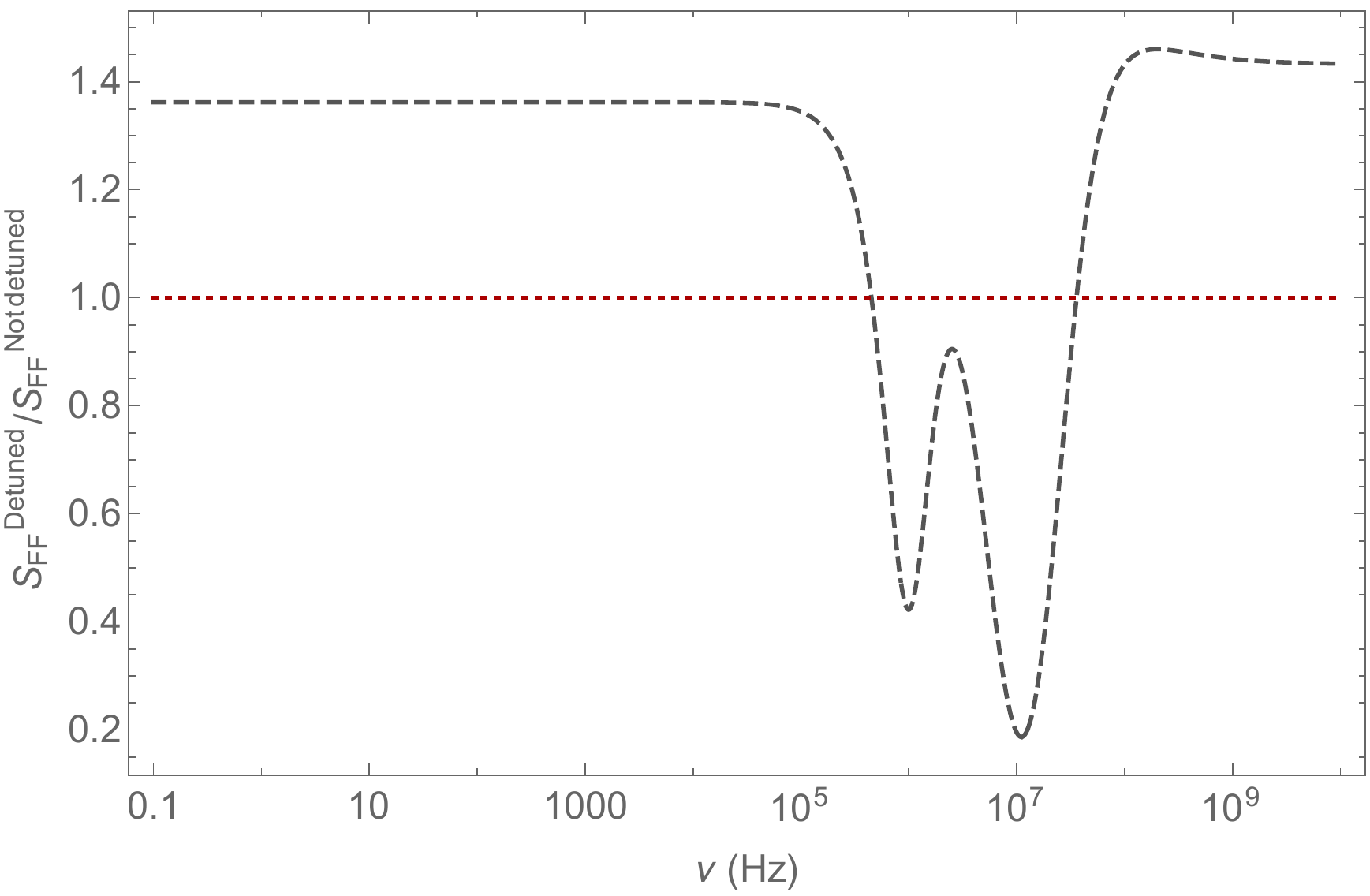}
\caption{Comparison between noise power spectral densities in a single-sided cavity with detuning and without detuning \eqref{noisepsdsingle}. Top: We show that the features of the noise PSD curves are similar when optimized. Here the detector parameters are the same as in figure \ref{figure-noisecontribs}, and we take detuning $\Delta \approx \kappa$. Bottom: We plot the ratio in between the noise PSD of detuned and non-detuned scenarios. Note that there is a frequency regime where the detuned noise PSD gives us lower noise than the non-detuned case. }
\label{figure-appendix}
\end{figure}

 We can solve for these equations of motion and obtain the following expression for the  amplitude and phase quadratures in the cavity :
\begin{equation}
    \begin{split}
    \begin{pmatrix}
     X  \\
     Y 
\end{pmatrix}  =
 \frac{1}{f } \begin{pmatrix}
 \frac{\kappa}{\chi_c \chi_m} & -\frac{\sqrt{\kappa}\Delta}{\chi_m}  & G \Delta  \\
\sqrt{\kappa} \left[G^2 \hbar+\frac{\Delta} {\chi_m}\right] & \frac{\kappa}{\chi_c \chi_m} & -\frac{G \sqrt{\kappa}}{\chi_c}  
\end{pmatrix} \begin{pmatrix}
X_{\rm in} \\ Y_{\rm in} \\ F_{\rm in}
\end{pmatrix}
\end{split}
\end{equation}
where 
\be
f=\frac{1}{{G^2 \hbar \Delta +\chi_m^{-1} (\kappa \chi_c^{-2} + \Delta^2 )}}.
\ee
Then we follow the input-output relations given in Eq. \eqref{input-output-single} to find the output quadratures. Since $\Delta \neq 0$ leads to mixing of the optical quadratures, the force signal $F_{in}$ is now imprinted on both $X_{\rm out}$ and $Y_{\rm out}$. We can choose an optimal quadrature to measure by linearly combining these so that the signal is in a single quadrature. Let's define
\begin{equation}
\begin{split}
a &= \frac{-G \Delta \sqrt{\kappa}  }{{G^2 \hbar \Delta +\chi_m^{-1} (\kappa \chi_c^{-2} + \Delta^2 )}}\\
b &= \frac{ G \kappa \chi_c^{-1} }{{G^2 \hbar \Delta +\chi_m^{-1} (\kappa \chi_c^{-2} + \Delta^2 )}}.
\end{split}
\end{equation}
Naively, we would like to consider the quadratures
\begin{align}
\begin{split}
    Q_{\rm out} & = a X_{\rm out} + b Y_{\rm out} \\
    P_{\rm out} & = b X_{\rm out} - a Y_{\rm out}.
\end{split}
\label{optimal-quadrature}
\end{align}
The $P$ quadrature here has no dependence on $F_{\text{in}}$, so all of the signal is encoded in the $Q$ quadrature. We would then like to monitor the $Q$ quadrature, but here the coefficients $a,b$ are functions of frequency $\omega$, so constructing the appropriate filter would be exceedingly difficult. Instead, we can just evaluate these coefficients at our frequency of interest $\omega_{\rm sig} \sim 1/\tau$. Using $a(\omega_{\rm sig}),b(\omega_{\rm sig})$, we define the observed quadrature
\be
Q_{\rm meas} = a(\omega_{\rm sig}) X_{\rm out} +  b(\omega_{\rm sig}) Y_{\rm out}.
\ee
To convert this measured output to a force, we divide through by the appropriate coefficient,
\be
F_E = \frac{Q_{\rm meas}}{a^2+b^2},
\ee
cf. equation \eqref{FE-single}. With these choices we can find the force noise PSD. Finally, we still have the freedom to optimize the optomechanical coupling $G$. As before, we can find  the optimized optomechanical coupling strength $G_{\rm opt}$ by minimizing the contribution from the measurement-added part of the noise, at the frequencies of interest around $1/\tau$. Doing so, and using the result in the noise PSD, we obtain the optimized noise PSD for the slightly detuned single sided cavity. See figure \ref{figure-appendix} for a comparison of the resulting noise PSD with the noise PSD in the case of exact cavity resonance \eqref{noisepsdsingle}.

\subsection{Slight Detuning in  Double Ring Cavities}
Similar to the single-sided cavity example illustrated above, we can introduce detuning to both the cavities in the system and obtain the following equations of motion:
\begin{align}
\begin{split}
\label{EOMdoubledetune}
&\dot{X} = -\frac{\kappa}{2}X+ \sqrt{\kappa} X_{in}- \Delta Y\\
&\dot{Y} = -G x -\frac{\kappa}{2}Y+ \sqrt{\kappa} Y_{in} + \Delta X\\
&\dot{X^{\prime}} = -G^\prime x \sin \theta -\frac{\kappa^\prime}{2}X^{\prime}+ \sqrt{\kappa^\prime} X_{in}^{\prime}- \Delta^\prime Y^{\prime}\\
&\dot{Y^{\prime}} = G^\prime x \cos \theta -\frac{\kappa^\prime}{2}Y^{\prime}+ \sqrt{\kappa^\prime} Y_{in}^{\prime}+ \Delta^\prime X^{\prime}\\
&\dot{p} = -\hbar G X +\hbar G^\prime \left( X^{\prime} \cos \theta+  Y^{\prime} \sin \theta \right)- m\omega_m^2 x -  \gamma p +  F_{in}\\ 
&\dot{x} = \frac{p}{m}
\end{split}
\end{align}
where the optomechanical coupling strength of the second cavity is related to that of the first cavity as
\begin{equation}
    G^\prime = G \sqrt{1-L} \sqrt{\frac{\kappa^\prime}{\kappa}} \sqrt{\frac{\Delta^2 +\kappa^2/4}{{\Delta^\prime}^2 +{\kappa^\prime}^2/4} }
\end{equation}
and the phase $\theta$ is defined as,
\be
e^{i \theta} = \frac{-i \Delta -\kappa/2}{-i \Delta^\prime + \kappa^\prime/2} \sqrt{\frac{{\Delta^\prime}^2 +{\kappa^\prime}^2/4}{\Delta^2 +\kappa^2/4}}.
\ee
It is hard to get tractable analytical expressions for the double ring cavity given the complexity of the coupled equations of motion. So, we will first qualitatively discuss the effect of introducing detuning into both of the cavities and then demonstrate some numerical results.

In standard displacement sensing, the mechanical position $x(t)$ is imprinted onto the light. Measurement of the light then causes backaction on the mechanics. Here, to avoid this backaction, we have suggested instead that one wants to monitor the mechanical velocity $v(t)$, with no measurement of position. In our two-cavity system, the light picks up a total phase shift $\Delta \phi = \phi_1 + \phi_2$. The condition that we do \emph{not} measure position then says that
\be
\label{backactioncond}
\frac{d}{dx} (\phi_1 + \phi_2) = 0.
\ee
We will show that this condition can be achieved by satisfying a simple constraint on the two cavity detunings $\Delta,\Delta^\prime$ and couplings $g_0,g_0'$.

Following the phase shift formalism introduced in the section above, the first cavity picks up a phase proportional to,
\be
a_{\rm out} = e^{i \phi_1} a_{\rm in}.
\ee
Now with a second cavity, in the limit that we can ignore losses in the delay line and treat the mechanics as stationary, the output light of the second cavity similarly picks up a phase shift involving $x$. We find
\begin{equation}
  a_{\rm out}^\prime = e^{i \phi_2} a_{\rm in}^\prime = e^{i(\phi_1 + \phi_2)} a_{\rm in}.
  \label{total-phase}
 \end{equation}
The total phase shift $\phi_1 + \phi_2$ is written in terms of the two effective detunings $\Delta_{\rm eff}, \Delta'_{\rm eff}$.

\begin{figure}[t]
\includegraphics[scale=0.45]{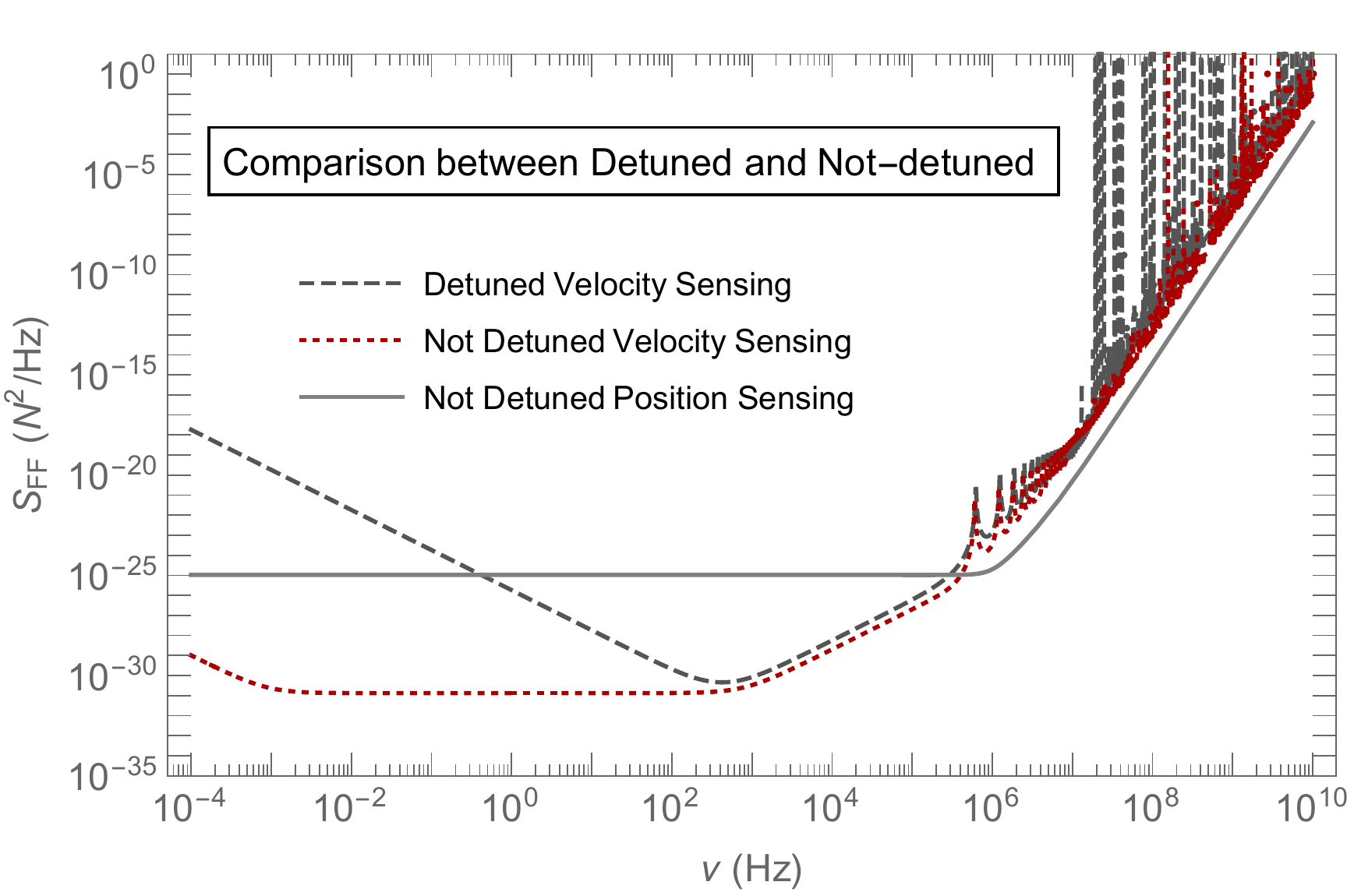}
\includegraphics[scale=0.43]{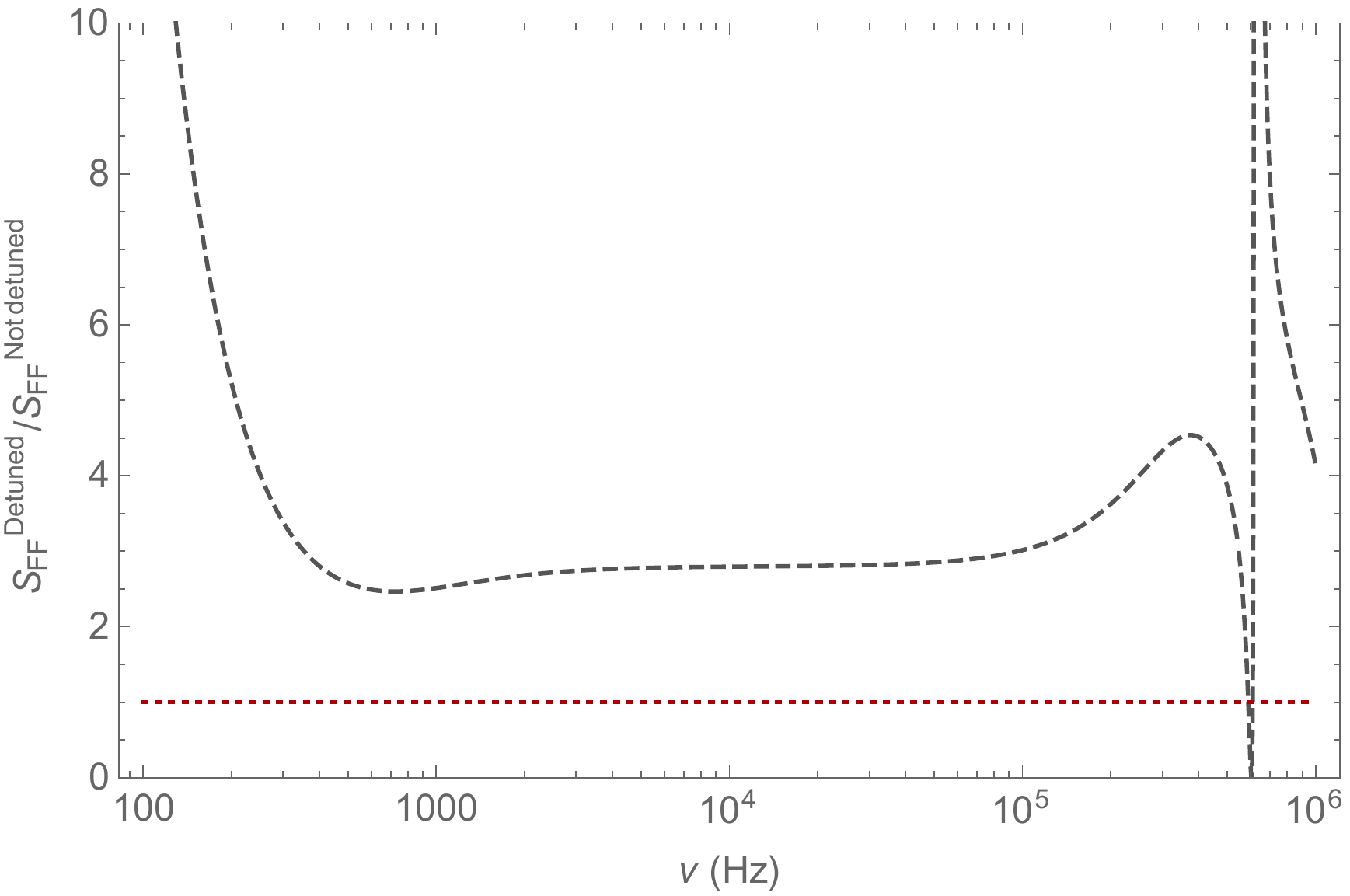}
\caption{Comparison between noise power spectral densities in a double ring cavity with detuning  and without detuning. Here the detector parameters are the same as in figure \ref{figure-noisecontribs}, and we take detuning $\Delta^\prime \approx \kappa \approx - \Delta $ assuming $g_0^\prime \sim -g_0$ and $\kappa^\prime \sim \kappa$. The optomechanical coupling strength is optimized for the non-detuned scenario as in \eqref{optimizedG} and the detuned case has been plotted for the same circulating power as in the non-detuned case. We see that in the region where we have optimized for backaction evasion, here for $10^4 \ {\rm Hz} < \nu < 10^6 \ {\rm Hz}$, the effects of detuning are minor once we impose our coupling condition \eqref{matchingcond}.The ratio plot at the bottom shows the comparison in this zoomed in region. }
\label{double-figure-appendix}
\end{figure} 

For a backaction evading measurement, our main goal is to eliminate the position dependence from this total phase. Mathematically this is just the statement of \eqref{backactioncond}. Using the above results for the phase shifts, at zero frequency of the mechanical oscillator, the phase matching condition reduces to
\begin{equation}
\begin{split}
\label{matchingcond}
\frac{{\Delta^{\prime}}^2+{\kappa^{\prime}}^2/4}{\Delta^2+\kappa^2/4} &= -\frac{g_0^\prime}{g_0} \frac{ \kappa^\prime}{\kappa } 
 \end{split}
\end{equation}
which relates the detunings and couplings in the two cavities. In the final expression we have assumed that for $g_0 \ll \Delta \sim \kappa $ : $\Delta^2_{\rm eff} \sim \Delta^2$ (and similarly ${\Delta^\prime}^2_{\rm eff} \sim {\Delta^\prime}^2$).  If this condition is satisfied, the measurement will evade mechanical backaction, which is straightforward for $g_0^\prime/g_0 <0$ as in our double cavity design and for $|\omega_c-\omega_c^\prime | \sim \kappa \sim \kappa^\prime$.

Thus we can choose detunings in both the cavities to satisfy the above ratio in order to evade the backaction noise.  After doing this, we solve the equations of motion and choose the optimal quadrature of the light coming out of the second cavity, similar to the  description given for the single-sided case \eqref{optimal-quadrature} to account for the total phase shift given in \eqref{total-phase}. This gives the noise PSD in the force estimator as before. This can be compared with the noise PSD in the case of zero detuning, cf. equation \eqref{optimizednoise}. The resulting formulas are too cumbersome to write down explicitly, but can be easily evaluated symbolically on a computer.

In figure \ref{double-figure-appendix}, we provide a numerical example comparing the noise in the detuned and non-detuned cases. The noise PSD in the non-detuned case is optimized with the optomechanical coupling strength given in \eqref{optimizedG}. For simplicity, we use the same circulating power inside the cavities for both scenarios which means use of higher amount of input laser power in the detuned case than the non-detuned case, although this choice could potentially be further optimized. From the figure, we see that in the frequency regime $\omega_m \ll \nu \lsim \kappa \sim \Delta$, the effects of relative cavity differences can be largely compensated for by optimizing the relative couplings and detunings, though we pay a price by having higher noise in the lower frequency regime of the PSD. As we are interested in signals in the radio band, we can see that our backaction-evasion strategy is highly robust to the presence of small mismatches in the two cavity parameters.

\bibliography{mybib}
\bibliographystyle{apsrev4-1}

\end{document}